\documentclass[prd,aps,10pt,a4paper,showkeys]{revtex4-2}
\usepackage[english]{babel}
\usepackage{amsmath,amssymb,amsmath}
\usepackage{dcolumn}
\usepackage{bm}
\usepackage[margin=1.5cm]{geometry}
\usepackage{subcaption}
\usepackage[pdftex]{graphicx}
\usepackage{mathtools}
\usepackage[T1]{fontenc}
\usepackage{microtype}
\usepackage[shortlabels]{enumitem}
\usepackage{tabularx}
\usepackage{multirow}
\usepackage[hidelinks,breaklinks]{hyperref}
\usepackage[capitalise]{cleveref}

\usepackage{xcolor}
\usepackage{orcidlink}
\usepackage{booktabs}

\definecolor{rose}{rgb}{1.0, 0.0, 0.5}

\hypersetup{
    colorlinks,
    linkcolor={rose},
    citecolor={green!60!black},
    urlcolor={cyan!50!blue}
}
\begin{document}

\title{\bf Constructions of Variable Modified Chaplygin-Jacobi and Chaplygin-Abel Gases from Geometrothermodynamics}

\author{Anamika Kotal \orcidlink{0009-0009-8426-3206}}
\email{kotalanamika31@gmail.com}
\author{Ujjal Debnath \orcidlink{0000-0002-2124-8908}}
\email{ujjaldebnath@gmail.com}

\affiliation{Department of Mathematics, Indian Institute of Engineering\\
Science and Technology, Shibpur, Howrah-711 103, India.}

%%%%%%%%%%%%%%%%%%%%%%%%%%%%%%%%%%%%%%%%%%%%%%%%%%%%%%%%%%%%%%%%%%%%%%%%%%%%%%%%%%%%%%%%%%%%%%%%%%%%%%%%%%%%%%%%%%%%%%%%%%%%%%%%%%%%%%%%%%%%%%%%%%%%%%%%%%%%%%%%%%%%%%%%%%%%%%%%%%%%%%%%%%%%%%%%%%

\begin{abstract}
{\large
Understanding the physical origin of the Universe's late-time accelerated expansion remains one of the central open problems in modern cosmology. Geometrothermodynamics (GTD) offers a compelling route to this problem, providing a geometric framework that unifies thermodynamics with contact and Riemannian geometry to encode the equilibrium properties of a physical system directly into its underlying geometric structure. In this work, we employ the GTD formalism to investigate cosmological models capable of explaining the observed accelerated expansion of the Universe. As an important outcome, our simple model exhibits dynamical behavior that provides a reasonable description of the late-time evolution of the Universe. We study the Chaplygin-Jacobi and Chaplygin-Abel gas models and their variable forms within geometrothermodynamics by deriving their thermodynamic relations, entropy, and equations of state. The free parameters of the models are constrained through a Markov Chain Monte Carlo (MCMC) analysis using several observational datasets, including cosmic chronometers combined with baryon acoustic oscillations (CC+BAO), DESI DR2 BAO data, DESI DR2 combined with the DES-SN5YR and Union3 Type Ia supernova compilations. The statistical performance of the models is assessed using the Akaike and Bayesian Information Criteria (AIC and BIC). We further examine their cosmological evolution through the deceleration parameter, the dark-energy equation-of-state parameter, and higher-order kinematic diagnostics, including the jerk and snap parameters. The perturbative stability of the models is also investigated through the squared sound speed. \\}\\

\end{abstract}

\maketitle

\newpage

\sloppy \tableofcontents

\section{\bf{Introduction}}

{

Observations of Type Ia supernovae (SNe Ia) \cite{SupernovaCosmologyProject:1997zqe,SupernovaCosmologyProject:1998vns,SupernovaSearchTeam:1998fmf} provided the first compelling evidence that the expansion of the Universe is accelerating. This conclusion has subsequently received support from several independent cosmological probes, including Baryon Acoustic Oscillations (BAO) \cite{Eisenstein:1997ik,SDSS:2005xqv}, Cosmic Microwave Background (CMB) measurements \cite{Bridle:2003yz}, and observations from the Wilkinson Microwave Anisotropy Probe (WMAP) \cite{WMAP:2003elm,WMAP:2008lyn} and the Planck mission \cite{LIGOScientific:2018glc}.
 To account for the observed accelerated expansion of the Universe, cosmological models introduce an exotic component known as dark energy (DE), characterized by a sufficiently negative pressure that produces an effective repulsive gravitational effect. Among the various proposed explanations for DE, the cosmological constant $\Lambda$, commonly interpreted as vacuum energy, remains the simplest framework and forms the basis of the standard $\Lambda$CDM cosmological model. Its contribution is estimated to constitute approximately 70\% of the total energy density of the present-day Universe \cite{Peebles:2002gy,padmanabhan2003gravity,Abdalla:2020ypg}. Although the concordance $\Lambda$CDM model provides an excellent description of a wide range of cosmological observations, the theoretical nature of the cosmological constant remains problematic, particularly due to the fine-tuning and cosmic coincidence issues \cite{Weinberg:1988cp}.
 Motivated by the theoretical challenges associated with the cosmological constant, a variety of dynamical dark energy scenarios have been proposed in which the energy density evolves with cosmic time \cite{Armendariz-Picon:2000ulo,Caldwell:1999ew,Elizalde:2004mq}. Among these possibilities, quintessence, generally described by a dynamical scalar field, has emerged as one of the widely investigated alternatives to the cosmological constant \cite{Tsujikawa:2013fta,Bamba:2012cp}. In addition to quintessence, several other scalar-field-based frameworks have been explored as possible explanations for the observed late-time acceleration, including different scalar field and tachyonic dark energy models \cite{Armendariz-Picon:2000nqq,sen2002tachyon,Feng:2004ad}.
 An alternative approach to explaining the late-time accelerated expansion is to modify the underlying gravitational theory rather than introduce a separate dark energy component. In such frameworks, modifications to the gravitational dynamics can effectively generate the acceleration of the Universe and thereby provide possible alternatives to dark energy \cite{li2013testing,Jacobson:2000xp,abdalla2005consistent,hovrava2009membranes,Barrow:2012qy}.
\\

The Chaplygin gas provides another phenomenological framework for describing the accelerated expansion without introducing dark energy as a separate component. This exotic fluid can reproduce the late-time acceleration of the Universe and, in its unified formulation, offers a description in which the dark matter and dark energy sectors emerge from a single cosmic fluid \cite{kamenshchik2000chaplygin,Gorini:2004by,Bilic:2001cg}.
 The Chaplygin gas framework was subsequently extended to the generalized Chaplygin gas (GCG), which retains the unified description of dark matter and dark energy within a single effective fluid \cite{Makler:2002jv,Bento:2002ps}. Further generalization led to the modified Chaplygin gas (MCG), whose cosmological evolution can interpolate between a radiation-dominated regime at early times and a $\Lambda$CDM-like behavior at late times \cite{Benaoum:2002zs,Debnath:2004cd}.
 Several further generalizations of the Chaplygin gas framework have been investigated, including the variable Chaplygin gas \cite{Guo:2005qy}, variable modified Chaplygin gas \cite{Debnath:2007bw}, and viscous Chaplygin gas \cite{Zhai:2005mu}. These extended formulations provide alternative phenomenological descriptions of dark energy and, in suitable settings, retain the possibility of interpreting dark matter and dark energy as manifestations of a single effective cosmic fluid. Observational data have been extensively employed to place constraints on the parameters governing Chaplygin gas models. In particular, measurements from Type Ia supernovae, the Cosmic Microwave Background (CMB), and WMAP have been used to constrain the parameters of the generalized Chaplygin gas (GCG) \cite{Bento:2002yx, Bento:2002ps}. Similar observational analyses based on CMB and WMAP data have also provided constraints on the parameters of the modified Chaplygin gas (MCG) \cite{Lu:2008zzb}. Furthermore, the behavior and allowed parameter ranges of the MCG have been investigated within various modified theories of gravity \cite{Chakraborty:2012mj,Paul:2012cxu,Debnath:2013cja,Debnath:2014saa,Ranjit:2014fra}.
 Villanueva \cite{villanueva2015generalized} formulated a generalized Chaplygin-Jacobi gas by constructing the Hubble parameter from a generalized Chaplygin scalar-field framework and employing Jacobi elliptic functions. Within this formulation, the corresponding inflationary observables and quantities were also derived. Inspired by this approach, we extend the analysis by introducing the equations of state for the modified Chaplygin-Jacobi gas and the modified Chaplygin-Abel gas.
\\

The geometric description of thermodynamic systems through metric structures has been widely studied. Different thermodynamic metrics are used to describe phase transitions through curvature singularities. The Fisher-Rao metric \cite{bhaduri2016classification}, arising from information theory, is expressed as the Hessian of entropy, while Ruppeiner's metric, based on thermodynamic fluctuation theory, is given by the negative Hessian of entropy \cite{Ruppeiner:1979bcp}. However, these metrics are not invariant under changes of thermodynamic representation or different thermodynamic potentials. For instance, in the energy representation \(U = U(E'^a)\), the metric is conformally related to the Weinhold metric, defined as the Hessian of energy \cite{Weinhold:1975xej}, with inverse temperature as the conformal factor. The Geometrothermodynamics (GTD) program provides a more general approach based on the invariance of its metrics under Legendre transformations. GTD has been applied to several systems, including ideal gases, van der Waals fluids \cite{Quevedo:2011np, Quevedo:2012jg}, and black holes. The invariance under changes of representation has also been investigated through the conformal structure of GTD metrics \cite{bravetti2013conformal}. In this work, we study the geometric properties of thermodynamic systems from a fully invariant perspective, considering invariance under both Legendre transformations and changes of representation. In this framework, dark energy (DE) is modeled as a geometrothermodynamic fluid (GTDF) with a negative equation of state, capable of counteracting gravitational attraction. This DE contribution becomes significant only at late cosmic times, emerging naturally after a transition redshift. The fluid’s pressure is proportional to its volume, so its negative pressure intensifies as the universe expands—an effect rooted in a constant thermodynamic interaction. Within geometrothermodynamics, such interaction is encoded in the constant curvature of the equilibrium manifold. This approach yields an evolving equation of state that asymptotically approaches a cosmological constant, offering a self-consistent late-time dynamics \cite{Bravetti:2013tcr}. However, a comprehensive investigation of Chaplygin-type realizations within the GTD framework is still lacking. Furthermore, most existing studies of Chaplygin-gas cosmology consider a limited number of fixed equations of state and are tested against only a restricted set of observational probes. Consequently, it remains important to examine how the inferred cosmological parameters and statistical performance depend on the specific functional form of the model and the choice of observational data, particularly in view of the differences among recent Type Ia supernova compilations such as DES-SN5YR, and Union3.

In this work, we address these issues by investigating two variable Chaplygin-type models within the GTD framework: the Variable Modified Chaplygin-Jacobi Gas (VMCJG) and the Variable Modified Chaplygin-Abel Gas (VMCAG). By deriving their fundamental thermodynamic relations and examining their cosmological evolution, we aim to answer the following key questions:

\begin{enumerate}
\item Can the VMCJG and VMCAG models provide a viable description of the late-time expansion history while remaining competitive with $\Lambda$CDM?

\item How do the two parametrizations differ in their predicted dark-energy dynamics, particularly in terms of the equation-of-state evolution, statefinder diagnostics, and higher-order kinematic parameters?

\item To what extent does the choice of supernova compilation, when combined with BAO data, affect the inferred values of $H_0$ and $\Omega_{m0}$, and can the additional flexibility of the Chaplygin-gas equation of state accommodate or reduce such compilation-dependent differences relative to $\Lambda$CDM?
\end{enumerate}
To address these questions, we perform a Markov Chain Monte Carlo likelihood analysis using four complementary data sets: CC+BAO, DESI DR2, DESI DR2+DES-SN5YR, and DESI DR2+Union3. For each combination, we compute the total $\chi^2$ and derive the posterior distributions of the model parameters. To assess the relative viability against $\Lambda$CDM, we evaluate the AIC, BIC, and DIC information criteria, which penalise model complexity while rewarding goodness of fit. Later, using the best-fitting values of the models' free
parameters, we study the evolutions of cosmological and
cosmographical parameters and check the classical stability of the models. The structure of this paper is as follows. In Sec.~\ref{rgeo}, we introduce the basic concepts of geometrothermodynamics (GTD) relevant to the present work. Sec.~\ref{cem} focuses on the construction of the entropy for the Modified Chaplygin-Jacobi Gas and Modified Chaplygin-Abel Gas within the GTD formalism, whereas Sec.~\ref{cevm} extends the analysis to their variable counterparts, namely the Variable Modified Chaplygin-Jacobi Gas and Variable Modified Chaplygin-Abel Gas. The fundamental field equations describing the cosmological dynamics are presented in Sec.~\ref{sec:field_equations}. In Sec.~\ref{sec:methodology}, we describe the methodology adopted for the observational analysis, including the MCMC likelihood approach and the data sets CC+BAO, DESI DR2, DESI DR2+DES-SN5YR, and DESI DR2+Union3. The observational results are presented in Sec.~\ref{find}, which is devoted to the complete cosmological analysis, including the estimation of the model parameters and their comparison using the AIC, BIC, and DIC criteria, while Sec.~\ref{cp} is devoted to the complete cosmological analysis. Finally, our conclusions are given in Sec.~\ref{con}.}

\section{\bf{Review of Geometrothermodynamics}}\label{rgeo}

In 2007, H. Quevedo \cite{Quevedo:2006xk} has developed
geometrothermodynamics (GTD) formalism to describe the properties
of thermodynamic systems by using differential geometry. In
classical equilibrium thermodynamics \cite{callen1960introduction}, the
thermodynamic system consists of $n$ extensive thermodynamic
variables $E^{a}~(a = 1, ..., n)$, their corresponding dual
intensive thermodynamic variables $I^{a}$ and the thermodynamic
potential $\Phi$. So define the phase space as the
$(2n+1)$-dimensional differential manifold ${\cal T}$  with a set
of coordinates $\{ Z^{A} \}=\{ \Phi,E^{a},I^{a} \}$ where
$A=1,2,...(2n+1)$. This $n$ determines the number of thermodynamic
degrees of freedom. The most general metric for GTD is
\begin{equation}
G=\Theta^{2}+\Lambda(E_{a}I_{a}dE^{a}dI^{a})
\end{equation}
where the fundamental Gibbs-1 form is
\begin{equation}
\Theta=d\Phi-I_{a}dE^{a}
\end{equation}
and $\Lambda\in \mathbb{R}$ is constant. The metric $G$ is
invariant under the Legendre transformations.

The $n$-dimensional equilibrium submanifold ${\cal E}\subset {\cal
T}$ can be defined by the smooth map $\varphi: {\cal E}
\rightarrow {\cal T}$ by $\varphi: \{ E^{a} \} \rightarrow  \{
\Phi(E^{a}), E^{a}, I^{a}(E^{a}) \}$ so that the condition
$\varphi^{*}(\Theta)=0$ (i.e., $d\Phi=I_{a}dE^{a}$ or,
$I_{a}=\frac{\partial \Phi}{\partial E^{a}}$) can be satisfied
where $\varphi^{*}$ is the pullback of $\varphi$.

The manifold ${\cal E}$ is naturally equipped with the induced
thermodynamic metric $g$, given by
\begin{equation}\label{g}
g=\varphi^{*}(G)=\Lambda\left(E_{a}\frac{\partial \Phi}{\partial
E^{a}}\right) \frac{\partial^{2}\Phi }{\partial E^{b} \partial
E^{c} } \delta^{ab} dE^{a}dE^{c}
\end{equation}

The equilibrium manifold ${\cal E}$ determines an extremal surface
in ${\cal T}$, so using variational principle, the variation of
the volume element is identically zero, i.e., $\delta \int_{{\cal
E}}\sqrt{det(g)}~d^{n}E=0$. Since the induced metric $g$ depends
on $Z^{A}$, so this variation derives the Nambu-Goto system of
differential equations
\begin{equation}
\Box
Z^{A}=\frac{1}{\sqrt{det(g)}}\left(\sqrt{det(g)}~g^{ab}Z^{A}_{,~a}
\right)_{,~b}+\Gamma^{A}_{BC}Z^{B}_{,~b}~Z^{C}_{,~c}~g^{bc}=0
\end{equation}
where $\Box$ denotes the d'Alembert operator and $\Gamma$ denotes
the Christoffel symbol for the metric $G_{AB}$. From the above
variation, one can obtain  a set of differential equations for the
thermodynamic potential $\Phi$ whose solutions can be expressed as
functions of the extensive variables $\Phi=\Phi(E^{a})$ which is
known as fundamental equations.

\section{\bf{Constructions of Entropy for Modified Chaplygin-Jacobi Gas and
Modified Chaplygin-Abel Gas using Geometrothermodynamics}}\label{cem}

We consider the entropy $S=S(U,V)$ as the thermodynamical
potential i.e., $\Phi=S$ and $U$, $V$ as the extensive variables
i.e., $E^{a}=\{U,V \}$. So $Z^{A}=\{S,U,V,\frac{1}{T},\frac{P}{T}
\}$. The fundamental Gibbs-1 form is
\begin{equation}
\Theta_{S}=dS-\frac{1}{T}dU-\frac{P}{T}dV
\end{equation}
For equilibrium state, the condition
$\varphi^{*}_{S}(\Theta_{S})=0$ reduces to the first law of
thermodynamics
\begin{equation}\label{dS}
dS=\frac{1}{T}dU+\frac{P}{T}dV
\end{equation}
such that
\begin{equation}\label{dS1}
\frac{\partial S}{\partial U}=\frac{1}{T}~,~\frac{\partial
S}{\partial V}=\frac{P}{T}
\end{equation}
In 2015, Villanueva \cite{villanueva2015generalized} has formulated the generalized
Chaplygin-Jacobi gas using Jacobi's elliptic function.
Subsequently, using Jacobi's elliptic function, we have derived
modified Chaplygin-Jacobi (MCJG) gas equation of state
\begin{equation}\label{MCJG}
P=[(2k-1)(1+A)-1]\rho-\frac{kB}{\rho^{\alpha}}+\frac{(1-k)(1+A)^{2}}{B}~\rho^{2+\alpha}
\end{equation}
where $A(>0),~B(>0),~\alpha\in(0,1)$ are constants and $k\in[0,1]$
is the elliptic modulus. Also using Abel's elliptic functions, we
have formulated modified Chaplygin-Abel gas (MCAG) equation of
state
\begin{equation}\label{MCAG}
P=[(e^{2}+2c^{2})(1+A)-1]\rho-\frac{c^{2}B}{\rho^{\alpha}}-\frac{(e^{2}+c^{2})(1+A)^{2}}{B}~\rho^{2+\alpha}
\end{equation}
where $c,e\in \mathbb{R}$. The above MCJG and MCAG can be
expressed in the form
\begin{equation}\label{MCJAG}
P=A_{1}~\rho-\frac{B_{1}}{\rho^{\alpha}}+C_{1}~\rho^{2+\alpha}
\end{equation}
where $A_{1}=(2k-1)(1+A)-1$, $B_{1}=kB$,
$C_{1}=\frac{(1-k)(1+A)^{2}}{B}$ for MCJG and
$A_{1}=(e^{2}+2c^{2})(1+A)-1$, $B_{1}=c^{2}B$,
$C_{1}=-\frac{(e^{2}+c^{2})(1+A)^{2}}{B}$ for MCAG. Since energy
density $\rho=U/V$, so the equation (\ref{MCJAG}) reduces to
\begin{equation}\label{MCJAG1}
P=\frac{A_{1}U^{1+\alpha}V^{1+\alpha}-B_{1}V^{2+2\alpha}+C_{1}U^{2+2\alpha}}{U^{\alpha}V^{2+2\alpha}}
\end{equation}
So equation (\ref{dS}) transforms to
\begin{equation}\label{dS2}
dS=\frac{1}{T}\left[MdU+NdV\right],~M=1,~N=P
\end{equation}
Since left hand side is the exact differential, so right hand side
should be the exact differential. Also the coefficients of $dU$
and $dV$ are homogeneous functions of $U$ and $V$. So $1/T$ is the
integrating factor and that is equal to $1/(MU+NV)$. So the
expression of temperature is
\begin{equation}
T=\frac{U^{\alpha}V^{2+2\alpha}}{(1+A_{1})U^{1+\alpha}V^{1+\alpha}-B_{1}V^{2+2\alpha}+C_{1}U^{2+2\alpha}}
\end{equation}
Putting the expression of $T$ in (\ref{dS2}) and integrating, we
obtain the expression of entropy as in the following form
\begin{equation}\label{S1}
S=S_{0}+\ln V+\frac{1}{2(1+\alpha)B_{1}D_{1}}\ln
\left[\frac{V^{1+\alpha}-\left(\frac{1+A_{1}}{2B_{1}}+D_{1}\right)U^{1+\alpha}
}{V^{1+\alpha}-\left(\frac{1+A_{1}}{2B_{1}}-D_{1}\right)U^{1+\alpha}}
\right]
\end{equation}
where $S_{0}$ is integration constant and
$D_{1}=\frac{(1+A_{1})^{2}+4B_{1}^{2}C_{1}}{4B_{1}^{2}}$.

\section{\bf{Variable Modified Chaplygin-Jacobi Gas and
Variable Modified Chaplygin-Abel Gas from Geometrothermodynamics}}\label{cevm}

Now consider the form of entropy
\begin{equation}\label{S2}
S=S_{0}+c_{1}\ln V+c_{2}\ln
\left[\frac{c_{3}U^{1+\alpha}+V^{1+\beta}}{c_{4}U^{1+\alpha}+V^{1+\beta}}
\right]
\end{equation}
where $c_{i}~(i=1,2,3,4)$ are constants. The induced metric
(\ref{g}) in the space of equilibrium states is given by
\begin{equation}
g=g_{UU}dU^{2}+2g_{UV}dUdV+g_{VV}dV^{2}
\end{equation}
where
\begin{eqnarray}
g_{UU}&&=\Lambda U\frac{\partial S}{\partial U}\frac{\partial^{2}
S}{\partial U^{2}} \nonumber
\\
&&=-\frac{\Lambda
c_{2}^{2}(c_{3}-c_{4})^{2}(1+\alpha)^{2}U^{2\alpha}V^{2+2\beta}}
{\left(c_{3}U^{1+\alpha}+V^{1+\beta}\right)^{3}\left(c_{4}U^{1+\alpha}+V^{1+\beta}\right)^{3}
} \nonumber
\\
&&\times\left\{
c_{3}U^{1+\alpha}\left(c_{4}(2+\alpha)U^{1+\alpha}+V^{1+\beta}\right)+V^{1+\beta}
\left(c_{4}U^{1+\alpha}-\alpha V^{1+\beta}\right)\right\},
\end{eqnarray}
\begin{eqnarray}
g_{UV}&&=\frac{\Lambda }{2} \left(U\frac{\partial S}{\partial
U}+V\frac{\partial S}{\partial V} \right) \frac{\partial^{2}
S}{\partial U \partial V} \nonumber
\\
&&=\frac{\Lambda
c_{2}(c_{3}-c_{4})(1+\alpha)(1+\beta)U^{\alpha}V^{\beta}\left(c_{3}c_{4}U^{2+2\alpha}-V^{2+2\beta}\right)}
{2\left(c_{3}U^{1+\alpha}+V^{1+\beta}\right)^{3}\left(c_{4}U^{1+\alpha}+V^{1+\beta}\right)^{3}
} \nonumber
\\
&&\times\left[c_{1}\left(c_{3}U^{1+\alpha}+V^{1+\beta}\right)\left(c_{4}U^{1+\alpha}+V^{1+\beta}\right)
+c_{2}(c_{3}-c_{4})(\alpha-\beta)U^{1+\alpha}V^{1+\beta} \right],
\end{eqnarray}
\begin{eqnarray}
g_{VV}&&=\Lambda V\frac{\partial S}{\partial V}\frac{\partial^{2}
S}{\partial V^{2}} \nonumber
\\
&&=-\Lambda \left\{
c_{1}-\frac{c_{2}(c_{3}-c_{4})(1+\beta)U^{1+\alpha}V^{1+\beta}}
{\left(c_{3}U^{1+\alpha}+V^{1+\beta}\right)\left(c_{4}U^{1+\alpha}+V^{1+\beta}\right)
} \right\} \nonumber
\\
&&\times\left[
\frac{c_{1}}{V^{2}}+\frac{c_{2}(c_{3}-c_{4})(1+\beta)U^{1+\alpha}V^{1+\beta}}
{\left(c_{3}U^{1+\alpha}+V^{1+\beta}\right)^{2}\left(c_{4}U^{1+\alpha}+V^{1+\beta}\right)^{2}
} \right. \nonumber
\\
&&\times\left.\left\{ c_{3}U^{1+\alpha}\left(c_{4}\beta
U^{1+\alpha}-V^{1+\beta}\right)-V^{1+\beta}
\left(c_{4}U^{1+\alpha}+(2+\beta)
V^{1+\beta}\right)\right\}\right]
\end{eqnarray}

For this induced metric, we can obtain the non-vanishing scalar
curvature. So the internal thermodynamic interaction is possible
for the Legendre invariant scalar curvature.

From equation (\ref{dS1}), we obtain
\begin{equation}\label{1T}
\frac{1}{T}=\frac{(1+\alpha)c_{2}(c_{3}-c_{4})U^{\alpha}V^{1+\beta}
}{\left(c_{3}U^{1+\alpha}+V^{1+\beta}\right)\left(c_{4}U^{1+\alpha}+V^{1+\beta}\right)}
\end{equation}
and
\begin{equation}\label{PT}
\frac{P}{T}=\frac{c_{1}}{V}-\frac{(1+\beta)c_{2}(c_{3}-c_{4})U^{1+\alpha}V^{\beta}
}{\left(c_{3}U^{1+\alpha}+V^{1+\beta}\right)\left(c_{4}U^{1+\alpha}+V^{1+\beta}\right)}
\end{equation}
From equations (\ref{1T}) and (\ref{PT}), we obtain
\begin{equation}\label{Prho}
P=\left[\frac{c_{1}(c_{3}+c_{4})}{(1+\alpha)c_{2}(c_{3}-c_{4})}
+\frac{1+\beta}{1+\alpha}
\right]\rho+\frac{c_{1}}{(1+\alpha)c_{2}(c_{3}-c_{4})}\frac{V^{\beta-\alpha}}{\rho^{\alpha}}
+\frac{c_{1}c_{3}c_{4}}{(1+\alpha)c_{2}(c_{3}-c_{4})}
V^{\alpha-\beta}\rho^{2+\alpha}
\end{equation}
If $\beta=\alpha$, the above EoS is similar with the MCJG and MCAG
equation of states (\ref{MCJG}) and (\ref{MCAG}). If we assume
$\frac{c_{1}}{c_{2}}=(1+\alpha)\sqrt{(2+A)^{2}-4k}$ and
$c_{3},c_{4}=-\frac{1}{2kB}\left[(2k-1)(1+A)-1 \pm
\sqrt{(2+A)^{2}-4k}\right]$, then the above expression
(\ref{Prho}) reduces to
\begin{equation}\label{VMCJG}
P=[(2k-1)(1+A)-1]\rho-\frac{kBV^{\beta-\alpha}}{\rho^{\alpha}}+\frac{(1-k)(1+A)^{2}}{BV^{\alpha-\beta}}~\rho^{2+\alpha}
\end{equation}
which is termed as ``Variable Modified Chaplygin-Jacobi Gas"
(VMCJG). Also if we assume
$\frac{c_{1}}{c_{2}}=(1+\alpha)\sqrt{(e^{2}(1+A)^{2}-1)^{2}-2c^{2}(1+A)}$
and $c_{3},c_{4}=-\frac{1}{2c^{2}B}\left[(e^{2}+2c^{2})(1+A)-1 \pm
\sqrt{(e^{2}(1+A)^{2}-1)^{2}-2c^{2}(1+A)} \right]$, then the above
expression (\ref{Prho}) reduces to
\begin{equation}\label{VMCAG}
P=[(e^{2}+2c^{2})(1+A)-1]\rho-\frac{c^{2}BV^{\beta-\alpha}}{\rho^{\alpha}}-\frac{(e^{2}+c^{2})(1+A)^{2}}{BV^{\beta-\alpha}}~\rho^{2+\alpha}
\end{equation}
which is termed as ``Variable Modified Chaplygin-Abel Gas"
(VMCAG). In the present geothermodynamic formulation, we assume the thermodynamic
volume to be proportional to the physical volume of the FLRW universe, $V\propto a^-3$. This choice allows the thermodynamic quantities to be consistently expressed in terms of the cosmological scale factor and its
evolution.\\

\section{\bf{More General Equation of State from Geometrothermodynamics}}

We can extend the form of entropy (\ref{S2}), so the more general
form of entropy can be considered as
\begin{equation}\label{S3}
S=S_{0}+c_{0}\ln V+\sum_{i=1}^{n}c_{i}\ln
\left(d_{i}U^{1+\alpha}+V^{1+\beta} \right)
\end{equation}
where $c_{i}$'s and $d_{i}$'s are constants. From equation
(\ref{dS1}), we obtain
\begin{equation}\label{2T}
\frac{1}{T}=(1+\alpha)\sum_{i=1}^{n}
\frac{c_{i}d_{i}U^{\alpha}}{\left(d_{i}U^{1+\alpha}+V^{1+\beta}
\right)}
\end{equation}
and
\begin{equation}\label{2PT}
\frac{P}{T}=\frac{c_{0}}{V}+(1+\beta)\sum_{i=1}^{n}
\frac{c_{i}V^{\alpha}}{\left(d_{i}U^{1+\alpha}+V^{1+\beta}
\right)}
\end{equation}
which turn on the general equation of state (GEoS)
\begin{equation}
P=\left[c_{0}+(1+\beta)V^{\beta-\alpha}\sum_{i=1}^{n}
\frac{c_{i}}{\left(d_{i}\rho^{1+\alpha}+V^{\beta-\alpha} \right)}
\right]\left[(1+\alpha)\rho^{\alpha}\sum_{i=1}^{n}
\frac{c_{i}d_{i}}{\left(d_{i}\rho^{1+\alpha}+V^{\beta-\alpha}
\right)} \right]^{-1}
\end{equation}
This expression reduces to the VMCJG and VMCAG if $n=2$ and
$c_{2}=-c_{1}$. For $n=1$, we can get VMCG.

\section{FRW Model}\label{sec:field_equations}

We assume a spatially flat, homogeneous, and isotropic universe described by the
Friedmann--Lema\^itre--Robertson--Walker (FLRW) metric \cite{Chattopadhyay:2010ht},
\begin{equation}
    ds^2 = -dt^2 + a^2(t)\left[dr^2 + r^2
    \left(d\theta^2 + \sin^2\theta\,d\phi^2\right)\right],
    \label{eq:flrw_metric}
\end{equation}
where $a(t)$ denotes the scale factor. The spatial curvature is set to $k=0$, in agreement with the spatially flat geometry supported by current cosmological
observations.

The evolution of the cosmic background is described by the Einstein field equations,
\begin{equation}
    G_{\mu\nu} \equiv R_{\mu\nu}
    - \frac{1}{2}R\,g_{\mu\nu}
    = 8\pi G\,T_{\mu\nu},
    \label{eq:einstein_eqs}
\end{equation}
where $R_{\mu\nu}$ and $R$ represent the Ricci tensor and Ricci scalar,
respectively, while $T_{\mu\nu}$ denotes the total energy--momentum tensor.
We adopt natural units, $8\pi G=c=1$, throughout the analysis.

The cosmic medium is assumed to consist of two non-interacting components:
pressureless matter, including dark matter and baryons, with density $\rho_m$
and pressure $p_m=0$, and a dark-energy component characterized by density
$\rho_{DE}$ and pressure $p_{DE}$. The total energy--momentum tensor can therefore
be written as
\begin{equation}
    T_{\mu\nu}=T_{\mu\nu}^{(m)}+T_{\mu\nu}^{(DE)},
    \qquad
    T_{\mu\nu}^{(i)}
    =(\rho_i+p_i)u_\mu u_\nu+p_i g_{\mu\nu},
    \label{eq:total_tmn}
\end{equation}
where $u^\mu$ is the four-velocity of the fluid satisfying
$u^\mu u_\mu=-1$, with $i\in\{m,DE\}$.

For the FLRW background \cite{kotal2026logarithmic}, the Einstein equations lead to the Friedmann equation
\begin{equation}
    H^2\equiv\left(\frac{\dot a}{a}\right)^2
    =\frac{1}{3}\left(\rho_m+\rho_{DE}\right),
    \label{eq:friedmann1}
\end{equation}
and the acceleration equation
\begin{equation}
    \frac{\ddot a}{a}
    =-\frac{1}{6}\left[
    \rho_m+\rho_{DE}+3p_{DE}\right],
    \label{eq:friedmann2}
\end{equation}
where $H$ is the Hubble parameter and an overdot denotes differentiation with
respect to cosmic time. Equivalently, the evolution of the Hubble parameter may
be expressed as
\begin{equation}
    \dot H+H^2
    =-\frac{1}{6}\left(\rho_m+\rho_{DE}+3p_{DE}\right).
    \label{eq:accel_eq}
\end{equation}

For non-interacting matter and dark-energy components, the conservation equations
follow from the Bianchi identity, $\nabla^\mu G_{\mu\nu}=0$. The matter component
satisfies
\begin{equation}
    \dot{\rho}_m+3H\rho_m=0,
    \qquad
    \rho_m=\rho_{m0}a^{-3},
    \label{eq:matter_conservation}
\end{equation}
where $\rho_{m0}$ is the present-day matter density. The dark-energy component,
on the other hand, obeys
\begin{equation}
    \dot{\rho}_{DE}
    +3H\left(\rho_{DE}+p_{DE}\right)=0.
    \label{eq:de_conservation}
\end{equation}
The latter equation requires a specific relation between $p_{DE}$ and
$\rho_{DE}$ to determine the evolution of the dark-energy density. In the
following section, we introduce the modified Chaplygin-type equation of state
adopted in this work and derive the corresponding cosmological evolution.

\section{Methodology}
\label{sec:methodology}

\subsection{Bayesian Framework}
\label{subsec:bayesian}

The free parameters of the cosmological model are constrained within a Bayesian statistical framework \cite{Schwarz:1978tpv,Spiegelhalter:2002yvw}. For a parameter vector $\boldsymbol{\theta}$ and a given dataset $D$, Bayes' theorem can be expressed as
\begin{equation}\label{eq:bayes}
   \mathcal{P}(\boldsymbol{\theta}\,|\,\mathbf{d})
\propto
\mathcal{L}(\mathbf{d}\,|\,\boldsymbol{\theta})\,
\pi(\boldsymbol{\theta}),
\end{equation}
where $P(\boldsymbol{\theta}|d)$ denotes the posterior probability distribution, $\mathcal{L}(d|\boldsymbol{\theta})$ represents the likelihood of obtaining the observed data for a given set of model parameters, and $\pi(\boldsymbol{\theta})$ is the prior distribution.
Assuming Gaussian-distributed observational uncertainties, the likelihood can be written in terms of the corresponding log-likelihood as
\begin{equation}
    \ln \mathcal{L}(D\,|\,\boldsymbol{\theta}) = -\frac{1}{2}
    \left[ \Delta \mathbf{X}^{T}\, \mathbf{C}^{-1}\, \Delta \mathbf{X} \right],
    \label{eq:loglike}
\end{equation}
where
$\Delta \mathbf{X} = \mathbf{X}_{\rm obs} - \mathbf{X}_{\rm th}(\boldsymbol{\theta})$
denotes the residual vector between the observed quantities and their corresponding theoretical predictions. Here, $\mathbf{C}$ represents the covariance matrix of the observational data. For datasets for which only independent statistical uncertainties are available, $\mathbf{C}$ reduces to a diagonal matrix with elements $\sigma_i^2$.
\begin{table}[htbp]
\centering
\caption{Uniform (flat) prior ranges adopted for the MCMC parameter estimation of the VMCJG and VMCAG gas models.}
\label{tab:priors}
\begin{tabular}{llc}
\toprule
Model & Parameter & Prior range \\
\midrule

\multirow{4}{*}{VMCJG}
 & $H_0$        & $\mathcal{U}(66, 73)$ \\
 & $\Omega_{m0}$ & $\mathcal{U}(0.1, 0.5)$ \\
 & $\alpha$      & $\mathcal{U}(0, 0.6)$ \\
 & $\beta$       & $\mathcal{U}(0.01, 1)$ \\
\midrule

\multirow{4}{*}{VMCAG}
 & $H_0$        & $\mathcal{U}(60, 72)$ \\
 & $\Omega_{m0}$ & $\mathcal{U}(0.2,0.35)$ \\
 & $\alpha$      & $\mathcal{U}(0.2, 1)$ \\
 & $\beta$       & $\mathcal{U}(0.4, 2.6)$ \\

\bottomrule
\end{tabular}
\end{table}
We impose flat (uniform) priors on all free parameters, with the allowed ranges chosen according to physically reasonable considerations and summarized in Table~\ref{tab:priors}. For statistically independent datasets, the combined likelihood is obtained by adding their individual log-likelihoods:
\begin{equation}
    \ln \mathcal{L}_{\rm joint} = \sum_{i} \ln \mathcal{L}_i,
    \label{eq:joint_like}
\end{equation}
thereby enabling the joint constraints to incorporate complementary information from different cosmological observations.

\subsection{Markov Chain Monte Carlo Sampling}\label{subsec:mcmc}
To sample the posterior distribution described by Eq.~\ref{eq:bayes}, we employ the affine-invariant ensemble Markov Chain Monte Carlo (MCMC) algorithm implemented in the \texttt{emcee} package \cite{foreman2013emcee}. Unlike conventional single-chain Metropolis--Hastings approaches, the affine-invariant sampler utilizes an ensemble of $N_{\rm walkers}$ walkers that explore the parameter space simultaneously. The proposal for each walker is constructed using the positions of the remaining walkers, making the method relatively insensitive to the scale and correlations among the model parameters and reducing the need for manually tuned proposal distributions \cite{ kotal2026fractal, kotal2026parameter}.

For each observational dataset, the walkers are initialized in a compact region surrounding a physically plausible starting point, with their initial positions drawn from a restricted portion of the adopted prior volume. Each ensemble is then evolved for $N_{\rm iter}$ iterations. To remove the influence of the initial conditions, the first $N_{\rm burn}$ iterations are discarded as the burn-in phase. The remaining samples are subsequently thinned by a factor of $N_{\rm thin}$, where appropriate, to reduce the effects of autocorrelation. The resulting samples are then flattened to obtain the final posterior distribution used for parameter estimation.

The marginalized constraints on individual parameters and their correlations are represented through one- and two-dimensional posterior distributions. The quoted parameter values correspond to the median of the marginalized posterior distributions, while the associated uncertainties are obtained from the 16th and 84th percentiles, representing the 68\% credible intervals. The posterior distributions are visualized using triangle plots generated at the $1\sigma$ and $2\sigma$ confidence levels, corresponding to the $68\%$ and $95\%$ credible regions, respectively.\texttt{GetDist} \cite{lewis2025getdist}.

For each MCMC analysis, the best-fit parameter vector, $\boldsymbol{\theta}_{\rm best}$, is identified with the maximum a posteriori (MAP) sample, i.e., the sample having the largest posterior probability within the flattened chain. We adopt the MAP point rather than constructing a parameter vector from the individual marginalized medians because, in the presence of strong parameter degeneracies and correlations, the latter combination need not correspond to an actual point within the high-probability region of the multidimensional posterior distribution.

\subsection{Data Description}
\label{subsec:data}

To constrain the model and examine its consistency with observations, we consider several complementary late-time cosmological probes. The analyses are performed both for individual datasets and for suitable combinations of independent datasets.

\subsubsection{Cosmic Chronometers and Baryon Acoustic Oscillations (CC+BAO)}

We consider a compilation of Hubble parameter measurements, $H(z_i)\pm\sigma_{H,i}$, obtained primarily through the cosmic chronometer (CC) differential-age technique, supplemented by BAO-based measurements \cite{Moresco:2018xdr, Yu:2017iju, Sharov:2018yvz}. The data cover a broad redshift interval extending approximately over $0\lesssim z\lesssim2$. For these measurements, the covariance matrix is taken to be diagonal, and the corresponding chi-square statistic is given by
$\chi^2_{\rm CC+BAO} = \sum_i \left[ H_{\rm obs}(z_i)
- H_{\rm th}(z_i;\boldsymbol{\theta}) \right]^2 / \sigma_{H,i}^2$
The associated likelihood is then obtained from Eq.~\eqref{eq:loglike}.

\subsubsection{DESI DR2 Baryon Acoustic Oscillations}

We further employ the Data Release 2 (DR2) BAO measurements from the Dark Energy Spectroscopic Instrument (DESI). The dataset consists of 12 BAO measurements involving the quantities $D_M/r_d$, $D_H/r_d$, and/or $D_V/r_d$, distributed over seven effective redshift bins in the range $0.295\leq z\leq2.33$. The correlations among the measurements are accounted for through the corresponding Gaussian covariance matrix \cite{Wang:2015tua, Xu:2016grp, Beutler:2011hx, Ross:2014qpa}.

The relevant distance measures are defined as
\begin{equation}
D_H(z)=\frac{c}{H(z)},
\end{equation}
and $D_V =[z\,D_M^2 D_H]^{1/3}$
where $D_M$ denotes the comoving angular-diameter distance. The sound horizon at the drag epoch, $r_d$, is fixed to the fiducial value $r_d = 148.3\,{\rm Mpc}$.

\subsubsection{Type Ia Supernovae}

Two independent Type Ia supernova (SN Ia) compilations are considered in our analysis \cite{Magana:2017nfs, SupernovaCosmologyProject:2008ojh, Pan-STARRS1:2017jku}:

\begin{itemize}

\item \textbf{DES-SN5YR:} The Dark Energy Survey five-year sample contains 1829 SNe Ia covering a comparable redshift range. Since the complete systematic covariance matrix is not available in our adopted implementation, we employ a diagonal covariance matrix constructed from the reported statistical uncertainties, $\sigma_{\mu,i}$. This treatment is less comprehensive than that adopted for Pantheon+SH0ES and may therefore lead to a modest underestimation of the parameter uncertainties \cite{deSouza:2025rhv, Rodrigues:2025tfg, Plaza:2025gcv}.

\item \textbf{Union3:} We also consider the Union3 compilation, represented by 22 redshift-binned SN Ia distance-modulus measurements spanning approximately $0.05\lesssim z\lesssim2.26$. The corresponding $22\times22$ covariance matrix is incorporated in the analysis, and the absolute-magnitude nuisance parameter is marginalized following the standard procedure \cite{betoule2014improved, Singh:2018oiq}.
\end{itemize}

For all three SN Ia datasets, the theoretical distance modulus is calculated according to
$\mu(z) = 5\log_{10}[D_L(z)/{\rm Mpc}] + 25$, with the
luminosity distance $D_L(z) = (1+z)\,D_M(z)$.

\subsection{Model Comparison Statistics}
\label{subsec:modelcomp}

To quantify the statistical performance of the proposed model and compare it with the standard $\Lambda$CDM scenario for different observational combinations, we employ the Akaike Information Criterion (AIC), Bayesian Information Criterion (BIC), and Deviance Information Criterion (DIC) \cite{liddle2007information, burnham2004multimodel, burnham2011aic}. These quantities are defined as
\begin{align}
    {\rm AIC} &= -2\ln \mathcal{L}_{\rm max} + 2k, \label{eq:aic}\\
    {\rm BIC} &= -2\ln \mathcal{L}_{\rm max} + k\ln N, \label{eq:bic}\\
    {\rm DIC} &= \overline{D(\boldsymbol{\theta})} + p_D,
    \label{eq:dic}
\end{align}
where $\mathcal{L}_{\rm max}$ denotes the maximum likelihood, $k$ is the number of free parameters in the model, and $N$ represents the total number of observational data points.

For the DIC, the deviance is defined as
$D(\boldsymbol{\theta}) =
-2\ln\mathcal{L}(\boldsymbol{\theta})$, $\overline{D(\boldsymbol{\theta})}$
is its posterior mean over the MCMC chain, and $p_D = \overline{D(\boldsymbol{\theta})}
- D(\overline{\boldsymbol{\theta}})$ is the effective number of
parameters, penalizing model complexity, with $\overline{D(\boldsymbol{\theta})}$ representing the posterior mean of the deviance, which accounts for the complexity of the model.

For all three information criteria, smaller values indicate a statistically preferred model. To facilitate a relative comparison, we define $\Delta{\rm AIC}$, $\Delta{\rm BIC}$, and $\Delta{\rm DIC}$ with respect to the model having the minimum value of the corresponding criterion. In general, differences $\Delta{\rm AIC}$ or $\Delta{\rm BIC}<2$ indicate no substantial statistical distinction between the competing models, values in the range $2$--$6$ provide positive evidence against the model with the larger value, while differences exceeding $10$ are conventionally interpreted as strong evidence in favor of the model with the lower criterion value.

\begin{figure}[ht]
    \centering
    \begin{subfigure}[b]{0.48\textwidth}
        \includegraphics[width=\textwidth]{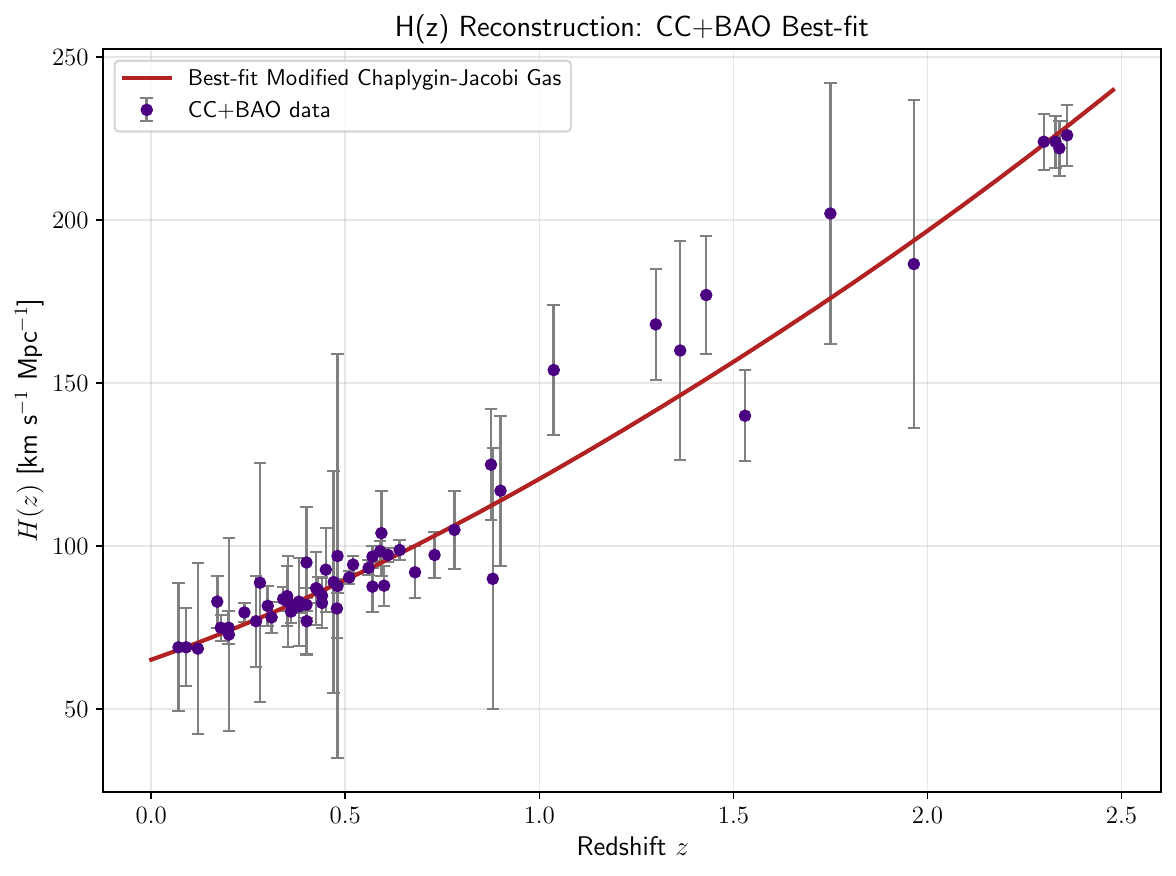}
        \caption{}
    \end{subfigure}
    \begin{subfigure}[b]{0.48\textwidth}
        \includegraphics[width=\textwidth]{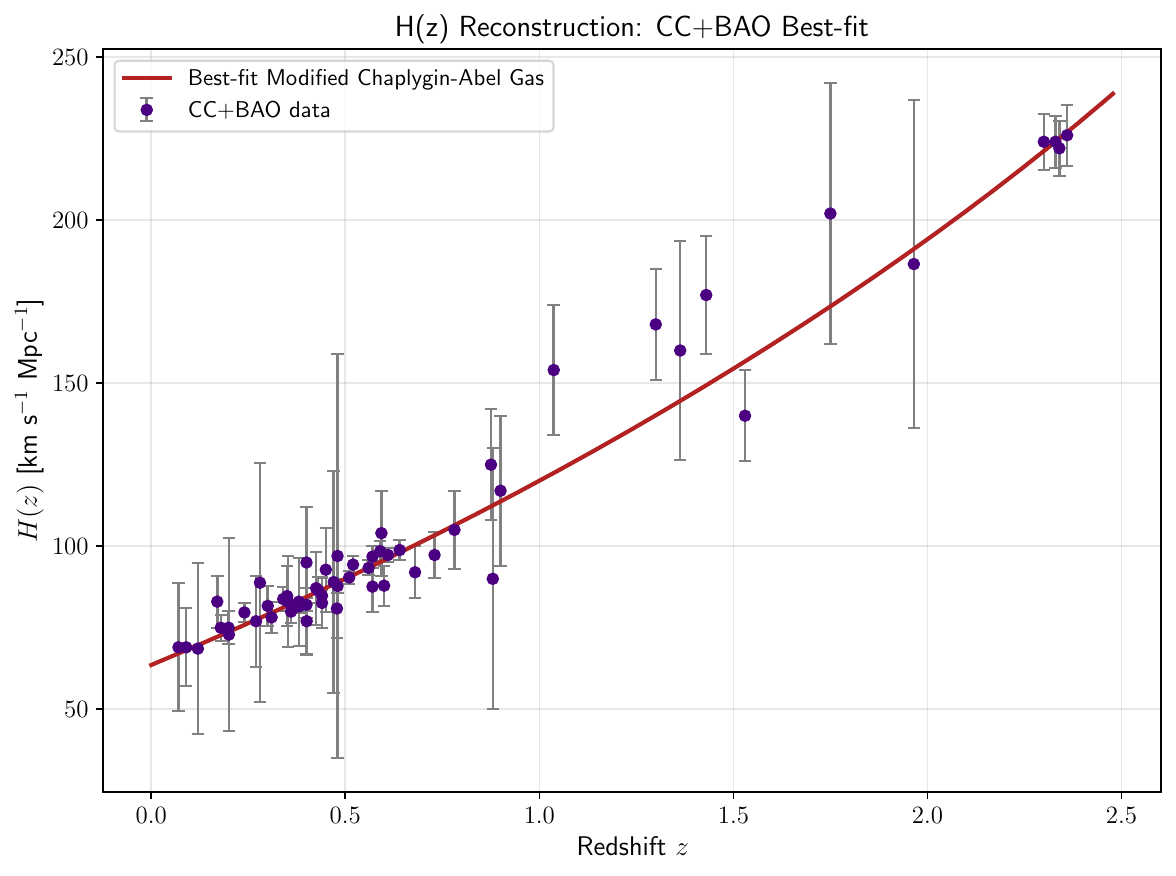}
        \caption{}
    \end{subfigure}
    \caption{The Hubble parameter $H(z)$ for the (a) Variable Modified Chaplygin Jacobi Gas and (b) Variable Modified Chaplygin Abel Gas models,
compared with the 57 CC+BAO observational data points.}
    \label{cc}
\end{figure}
\begin{figure}[ht]
    \centering
    \begin{subfigure}[b]{0.48\textwidth}
        \includegraphics[width=\textwidth]{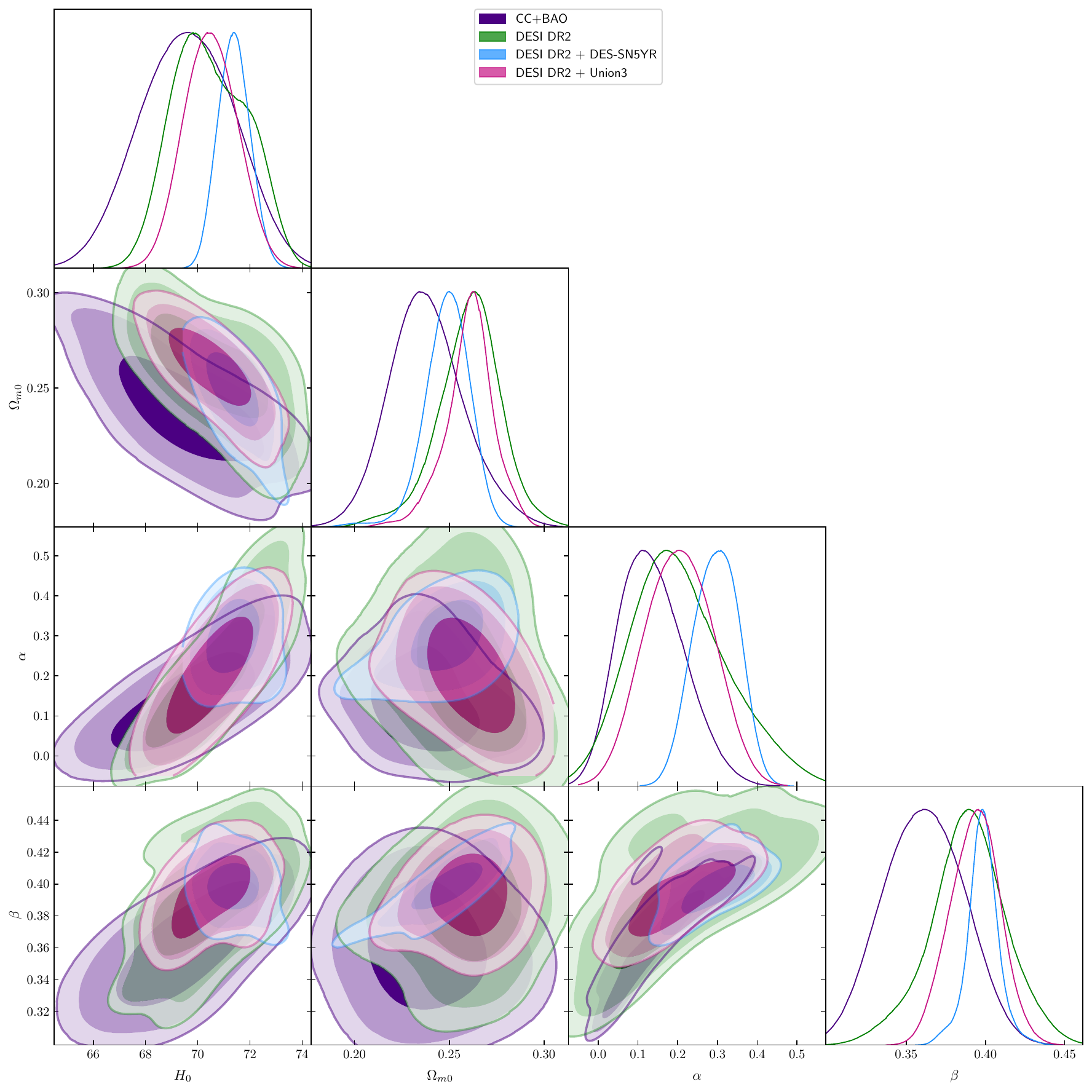}
        \caption{}
    \end{subfigure}
\begin{subfigure}[b]{0.48\textwidth}
\includegraphics[width=\textwidth]{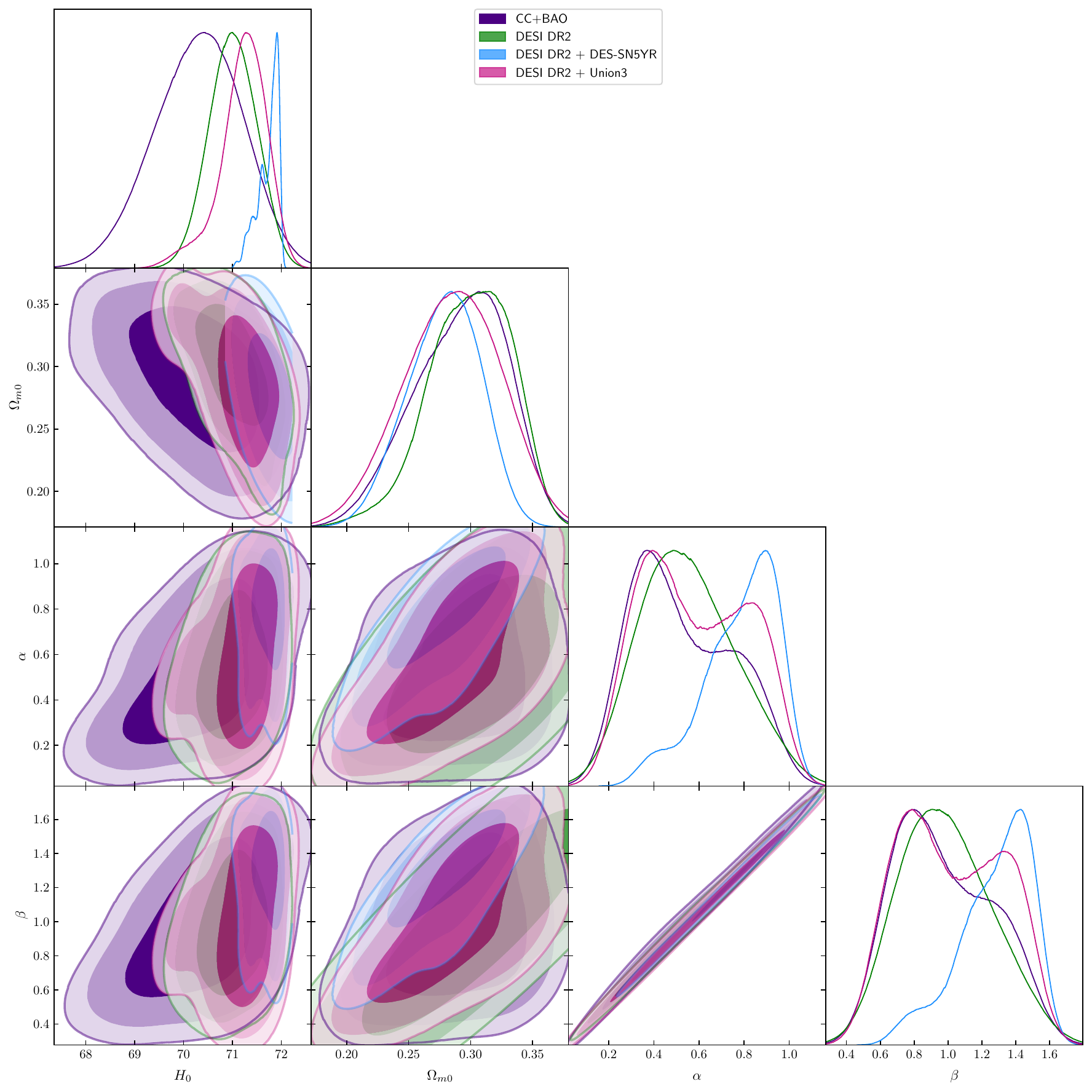}
\caption{}
    \end{subfigure}
    \begin{subfigure}[b]{0.48\textwidth}
\includegraphics[width=\textwidth]{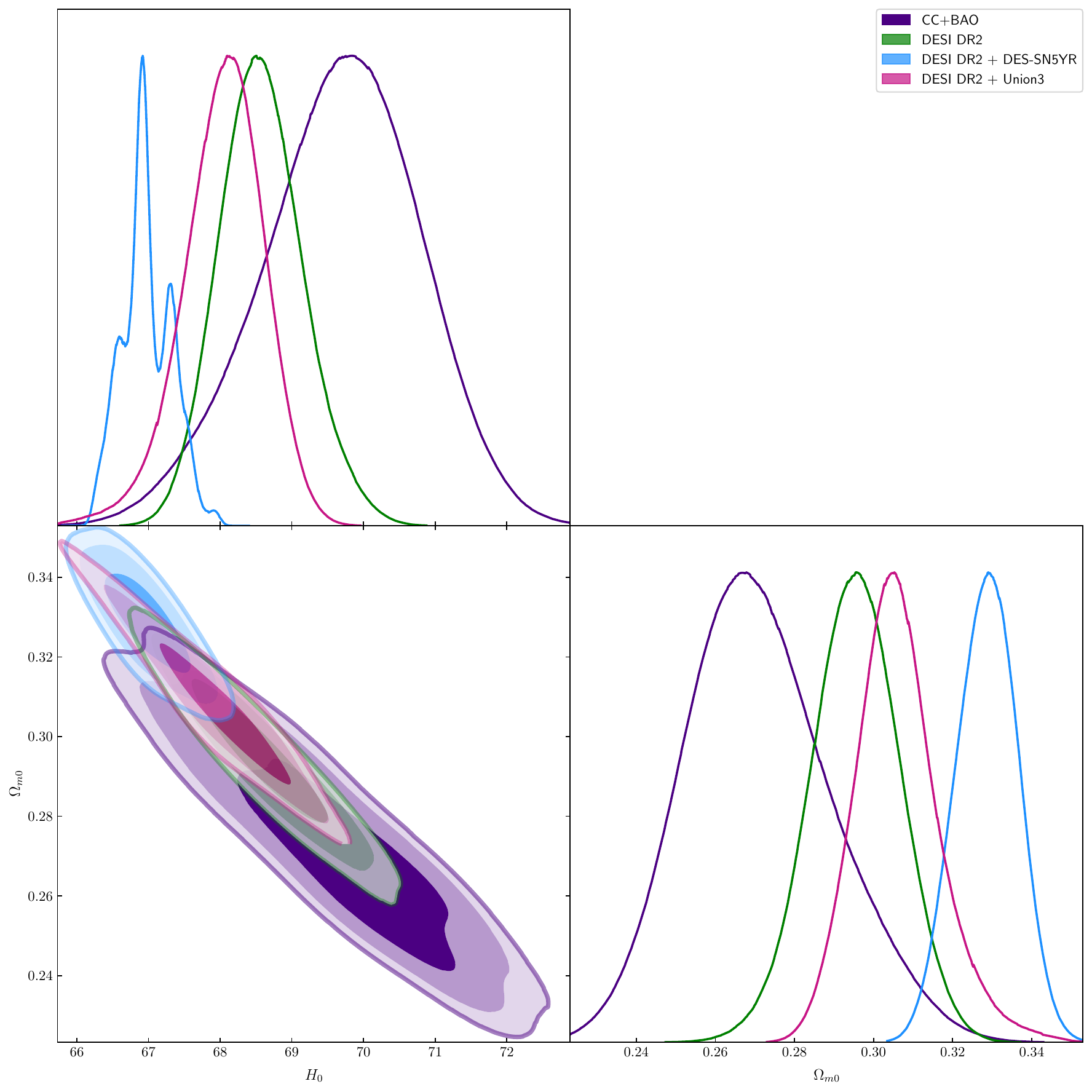}
\caption{}
    \end{subfigure}

    \caption{Marginalized one- and two-dimensional posterior distributions for the
free parameters of the (a)Variable Modified Chaplygin Jacobi Gas, (b)
Variable Modified Chaplygin--Abel Gas, (c) $\Lambda$CDM models.}
    \label{getd}
\end{figure}
\section{Observational Analysis}\label{find}
Both reconstructed $H(z)$ curves accurately track the CC+BAO data across the full redshift range (
$z<2.4$), remaining well within the observational uncertainties at all points, as seen from Fig.\ref{cc}. The VMCJG and VMCAG best-fit curves are visually indistinguishable from one another, indicating that the two parametrizations yield nearly identical background expansion histories despite their differing equations of state. This near-degeneracy at the level of $H(z)$ suggests that CC+BAO data alone cannot discriminate between the two models, and that additional probes such as Type Ia supernovae and DESI BAO distance measurements are needed to break this degeneracy.\\
From the getdist plots in Fig.\ref{getd},both models constrain $H_0$ to $\sim69$--$72$ km s$^{-1}$ Mpc$^{-1}$ and $\Omega_{m0}$ to $\sim0.24$--$0.30$, consistent with the standard $\Lambda$CDM scenario and lying between the Planck and SH0ES constraints. The posterior distributions also show the expected inverse trend between $H_0$ and $\Omega_{m0}$. For the Chaplygin-Jacobi model, $\alpha$ and $\beta$ are relatively well constrained, with $\alpha\sim0$-$0.45$ and $\beta\sim0.30$-$0.45$. In contrast, the $\alpha$-$\beta$ contours of the Chaplygin-Abel model are strongly elongated along a diagonal direction, indicating a pronounced parameter degeneracy. This feature is also reflected in the bimodal one-dimensional posterior distributions of $\alpha$ and $\beta$. Overall, the Chaplygin-Jacobi parametrization provides comparatively tighter and more well-defined constraints on the model parameters, indicating a better constrained parameter space than the Chaplygin-Abel case.
\begin{table*}[htbp]
\centering
\caption{Constraints on the cosmological parameters for the
$\Lambda$CDM, Chaplygin--Abel, and Chaplygin--Jacobi models. The quoted
values represent the posterior median with $1\sigma$ uncertainties.}
\label{tab:parameter_constraints}
\resizebox{\textwidth}{!}{%
\begin{tabular}{llcccc}
\toprule
Model & Dataset & $H_0$ (km s$^{-1}$ Mpc$^{-1}$)
& $\Omega_{m0}$ & $\alpha$ & $\beta$ \\
\hline

\multirow{5}{*}{$\Lambda$CDM}
& CC+BAO & $69.867_{-0.903}^{+0.826}$ & $0.268_{-0.018}^{+0.015}$ & -- & -- \\
& DESI DR2 & $68.551_{-0.493}^{+0.451}$ & $0.296_{-0.009}^{+0.008}$ & -- & -- \\
& DESI DR2 + DES-SN5YR & $66.943_{-0.491}^{+0.382}$ & $0.329_{-0.005}^{+0.010}$ & -- & -- \\
& DESI DR2 + Union3 & $68.041_{-0.555}^{+0.467}$ & $0.305_{-0.008}^{+0.011}$ & -- & -- \\
\hline

\multirow{5}{*}{VMCAG}
& CC+BAO
& $70.509^{+0.706}_{-1.253}$
& $0.299^{+0.039}_{-0.049}$
& $0.487^{+0.313}_{-0.174}$
& $0.927^{+0.407}_{-0.215}$ \\

& DESI DR2
& $71.055^{+0.600}_{-0.480}$
& $0.292^{+0.031}_{-0.053}$
& $0.457^{+0.294}_{-0.197}$
& $0.865^{+0.377}_{-0.259}$ \\

& DESI DR2 + DES-SN5YR
& $71.790^{+0.154}_{-0.286}$
& $0.284^{+0.026}_{-0.028}$
& $0.864^{+0.099}_{-0.199}$
& $1.377^{+0.136}_{-0.260}$ \\

& DESI DR2 + Union3
& $71.343^{+0.379}_{-0.633}$
& $0.288^{+0.041}_{-0.036}$
& $0.553^{+0.272}_{-0.211}$
& $0.999^{+0.342}_{-0.294}$ \\

\hline

\multirow{5}{*}{VMCJG}
& CC+BAO
& $69.675^{+1.765}_{-2.176}$
& $0.238^{+0.021}_{-0.026}$
& $0.090^{+0.130}_{-0.045}$
& $0.354^{+0.028}_{-0.027}$ \\

& DESI DR2
& $69.634^{+1.732}_{-1.132}$
& $0.264^{+0.014}_{-0.014}$
& $0.128^{+0.105}_{-0.088}$
& $0.378^{+0.021}_{-0.025}$ \\

& DESI DR2 + DES-SN5YR
& $71.396^{+0.636}_{-0.566}$
& $0.249^{+0.013}_{-0.012}$
& $0.281^{+0.035}_{-0.052}$
& $0.399^{+0.012}_{-0.010}$ \\

& DESI DR2 + Union3
& $70.553^{+0.717}_{-1.121}$
& $0.259^{+0.014}_{-0.012}$
& $0.200^{+0.065}_{-0.110}$
& $0.391^{+0.014}_{-0.023}$ \\

\bottomrule
\end{tabular}%
}
\end{table*}

\begin{table}[htbp]
\centering
\caption{Model comparison across CC+BAO, DESI DR2, DESI DR2 + DES-SN5YR, and DESI DR2 + Union3 for the VMCAG, VMCJG, and $\Lambda$CDM models. Reduced $\chi^2 \equiv -2\ln\mathcal{L}/(N-k)$, where $N$ is the number of data points and $k$ is the number of free parameters ($k=4$ for both Chaplygin variants, $k=2$ for $\Lambda$CDM. Boldface marks the preferred model (lowest AIC/BIC/DIC) within each dataset.}
\label{tab:model_comparison}
\small
\begin{tabular}{llccccc}
\toprule
Dataset & Model & $N$ & Reduced $\chi^2$ & AIC & BIC & DIC \\
\midrule
\multirow{3}{*}{CC+BAO}
 & VMCAG  & 57 & 0.573 & 38.39 & 46.56 & \textbf{34.91} \\
 & VMCJG & 57 & \textbf{0.552} & {37.24} & 45.42 & 36.00 \\
 & $\Lambda$CDM & 57 & 0.597 & \textbf{36.18} & \textbf{40.27} & {35.56}$^{\dagger}$ \\
\midrule
\multirow{3}{*}{DESI DR2}
 & VMCAG   & 12 & 0.800 & 14.40 & 16.34 & \textbf{10.81} \\
 & VMCJG & 12 & \textbf{0.720} & \textbf{13.76} & {15.70} & 11.88 \\
 & $\Lambda$CDM & 12 & 1.141 & 14.27 & \textbf{15.24} & 13.47 \\
\midrule
\multirow{3}{*}{DESI DR2 + DES-SN5YR}
 & VMCAG  & 1841 & 0.933 & 1721.87 & 1743.94 & 1720.98 \\
 & VMCJG & 1841 & \textbf{0.921} & \textbf{1700.03} & \textbf{1722.11} & \textbf{1700.34} \\
 & $\Lambda$CDM & 1841 & 0.933 & 1718.53 & 1729.57 & 1718.29 \\
\midrule
\multirow{3}{*}{DESI DR2 + Union3}
 & VMCAG   & 34 & 1.050 & 39.49 & 45.60 & 36.92 \\
 & VMCJG & 34 & \textbf{0.946} & \textbf{36.39} & \textbf{42.49} & \textbf{36.25} \\
 & $\Lambda$CDM & 34 & 1.246 & 42.63 & 45.68 & 42.09 \\
\bottomrule
\end{tabular}
\end{table}
% =====================================================================%

\begin{table}[htbp]
\centering
\caption{$\Delta$AIC, $\Delta$BIC, and $\Delta$DIC relative to the
$\Lambda$CDM reference model, for the VMCAG and VMCJG Gas models across all dataset combinations. Negative values indicate the model is preferred over $\Lambda$CDM($+\nu$) by that criterion; positive values indicate $\Lambda$CDM is preferred\cite{liddle2007information}.}
\label{tab:delta_lcdm}
\small
\begin{tabular}{llccc}
\toprule
Dataset & Model & $\Delta$AIC & $\Delta$BIC & $\Delta$DIC \\
\midrule
\multirow{3}{*}{CC+BAO}
 & VMCAG   & $+2.21$ & $+6.29$ & $-0.65$ \\
 & VMCJG & $+1.06$ & $+5.15$ & $+0.44$ \\
 & $\Lambda$CDM & $0.00$ & $0.00$ & $0.00$ \\
\midrule
\multirow{3}{*}{DESI DR2}
 & VMCAG   & $+0.13$ & $+1.10$ & $-2.66$ \\
 & VMCJG & $-0.51$ & $+0.46$ & $-1.59$ \\
 & $\Lambda$CDM & $0.00$ & $0.00$ & $0.00$ \\
\midrule
\multirow{3}{*}{DESI DR2 + DES-SN5YR}
 & VMCAG   & $+3.34$ & \textbf{$+14.37$} & $+2.69$ \\
 & VMCJG & \textbf{$-18.50$} & $-7.46$ & \textbf{$-17.95$} \\
 & $\Lambda$CDM & $0.00$ & $0.00$ & $0.00$ \\
\midrule
\multirow{3}{*}{DESI DR2 + Union3}
 & VMCAG   & $-3.14$ & $-0.08$ & $-5.17$ \\
 & VMCJG & $-6.24$ & $-3.19$ & $-5.84$ \\
 & $\Lambda$CDM & $0.00$ & $0.00$ & $0.00$ \\
\bottomrule
\end{tabular}
\end{table}
The best-fit parameters, presented in Table. \ref{tab:parameter_constraints} reveal a clear distinction between the Chaplygin models and $\Lambda$CDM. While the $\Lambda$CDM value of $H_0$ decreases with the inclusion of supernova data, from $69.87$ for CC+BAO to
$66.94\,\mathrm{km\,s^{-1}\,Mpc^{-1}}$ for DESI DR2+DES-SN5YR, both Chaplygin
models consistently favor higher values of $H_0\sim70$-$72\,
\mathrm{km\,s^{-1}\,Mpc^{-1}}$. The difference becomes particularly pronounced
for the SN-inclusive datasets, suggesting that the additional freedom in the
Chaplygin equations of state allows a higher expansion rate closer to local
measurements. A similar trend is found for $\Omega_{m0}$, with both Chaplygin
models yielding lower and more stable matter densities than $\Lambda$CDM; the
Jacobi model gives the lowest values, $\Omega_{m0}\sim0.237$-$0.264$. The
shape parameters also show a consistent hierarchy, with the Abel model favoring
larger $\alpha$ and $\beta$ values ($\alpha\sim0.46$-$0.86$,
$\beta\sim0.87$-$1.40$) than the Jacobi model
($\alpha\sim0.09$-$0.28$, $\beta\sim0.35$-$0.40$). Both parameters increase
notably when DES-SN5YR is included, indicating that supernova data play a key
role in driving the departure from $\Lambda$CDM, particularly for the Jacobi
parametrization, which is also strongly supported by the information criterion
analysis.\\

Table~\ref{tab:delta_lcdm} reports $\Delta\text{AIC}$, $\Delta\text{BIC}$, and
$\Delta\text{DIC}$, defined as $\text{IC}_{\text{model}} -\text{IC}_{\Lambda\text{CDM}}$; negative values therefore indicate that the
given Chaplygin-type model achieves a lower, more favorable information criterion
score than $\Lambda$CDM and is statistically preferred by that criterion, while positive values indicate $\Lambda$CDM is preferred. For CC+BAO, $\Lambda$CDM is mildly favored over both Chaplygin models by the AIC and BIC values, aside from a small preference for the Abel model by the DIC ($\Delta\text{DIC}=-0.65$). For DESI DR2 alone, no criterion reaches even positive evidence for $\Lambda$CDM($+\nu$); the DIC in fact mildly favors both Chaplygin models, though these fall short of strong evidence. The clearest result emerges for DESI DR2 combined with DES-SN5YR: the VMCJG model is very strongly preferred over $\Lambda$CDM, whereas the VMCAG model is disfavored relative to $\Lambda$CDM by all three criteria. For DESI DR2 combined with Union3, both Chaplygin models are preferred over $\Lambda$CDM: the VMCJG model shows strong
evidence via the AIC ($\Delta\text{AIC}=-6.24$) and positive evidence via the BIC and DIC, while the VMCAG model shows positive evidence via the AIC and DIC but no significant preference via the BIC ($\Delta\text{BIC}=-0.08$). Taken together, the VMCJG model is consistently preferred over, or statistically indistinguishable from, $\Lambda$CDM across all four dataset combinations, with a decisive
preference once supernova data are included, whereas the VMCAG model is comparable to or mildly preferred over $\Lambda$CDM for CC+BAO, DESI DR2, and Union3, but is disfavored specifically for the DES-SN5YR combination.

\section{Cosmographical Parameters}\label{cp}

{
In this section, we'll study the nature of the cosmological and
cosmographical parameters by taking the best-fit values of the
parameters of the VMCJG and VMCAG models \cite{Pan:2017ent}. \\

$\bullet{}$ EoS parameter:\\

The propagation of VMCJG and VMCAG for the evolution of the universe can be determined by the equation of state parameter $W$ \cite{Escamilla-Rivera:2019aol}.
\begin{figure}[ht]
    \centering
    \begin{subfigure}[b]{0.48\textwidth}
        \includegraphics[width=\textwidth]{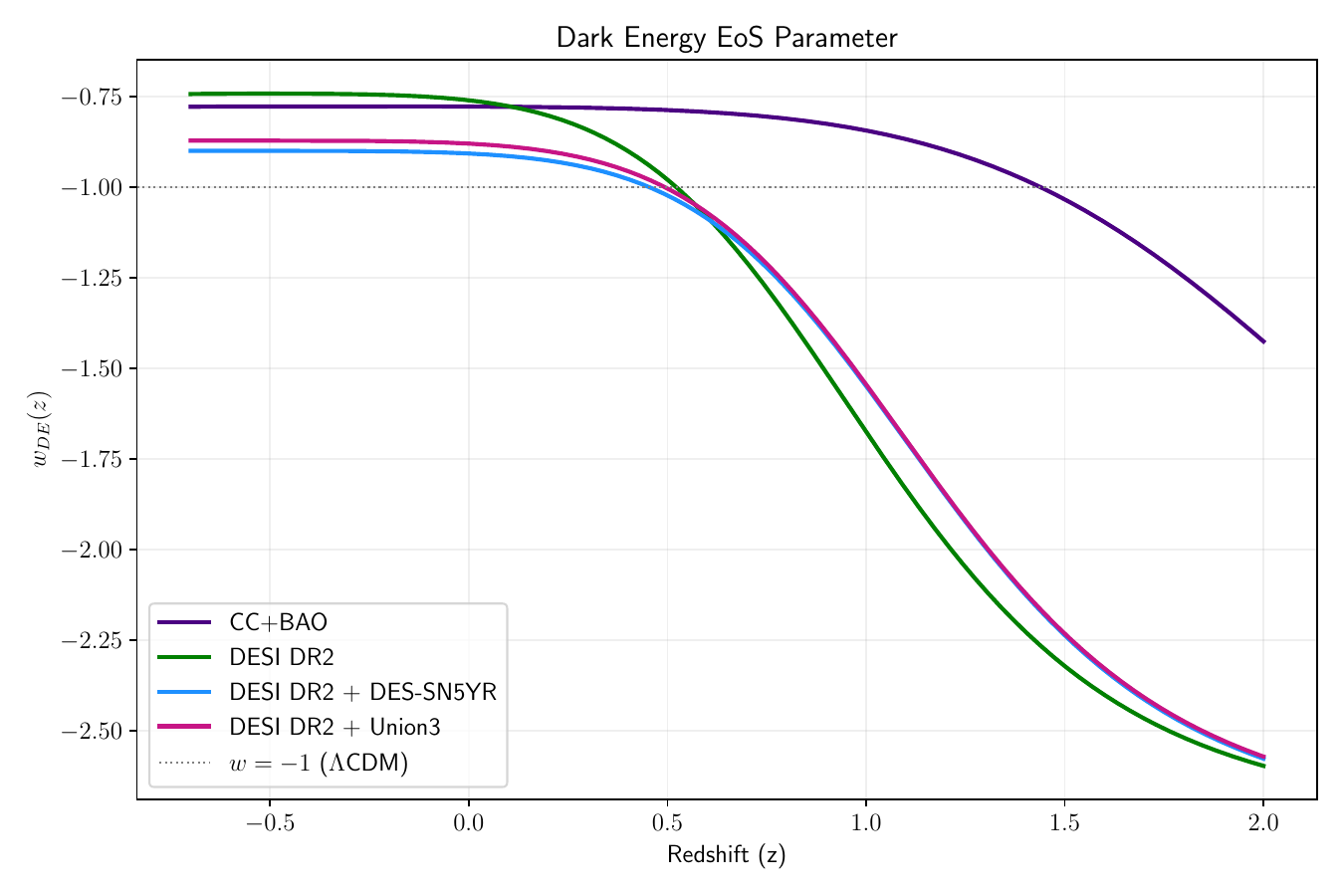}
        \caption{}
    \end{subfigure}
    \begin{subfigure}[b]{0.48\textwidth}
\includegraphics[width=\textwidth]{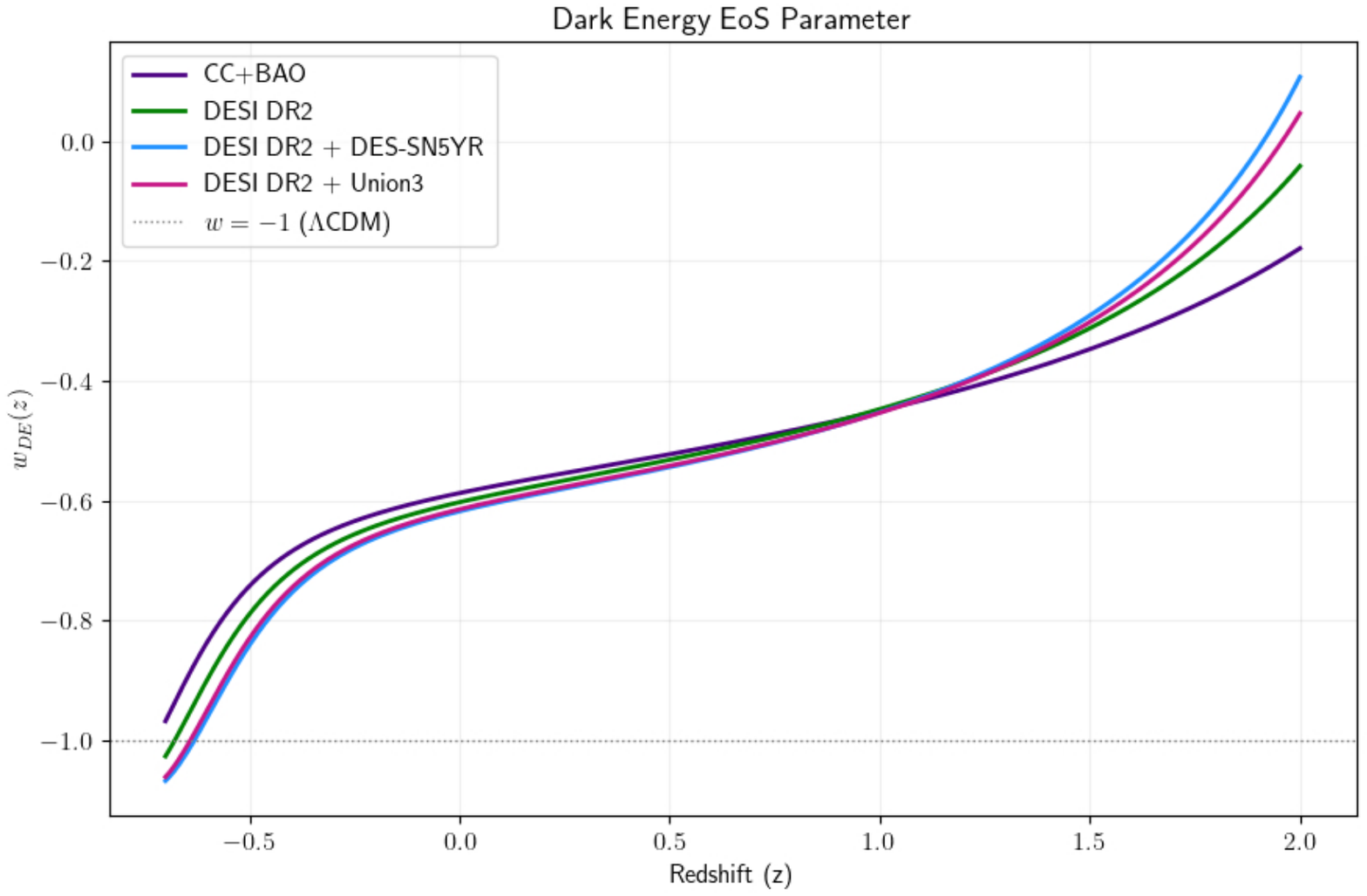}
        \caption{}
    \end{subfigure}
    \caption{Evolution of the EoS parameter $\omega(z)$ for the
(a) Variable Modified Chaplygin-Jacobi Gas (b) Variable Modified
Chaplygin-Abel Gas models for the considered observational datasets.}
    \label{eos}
\end{figure}

The two models exhibit a significant difference in the redshift evolution of the dark-energy equation-of-state parameter $w_{DE}(z)$, as plotted in Fig.\ref{eos}. Both models cross the phantom divide, $w=-1$, around $z\sim0.3$-$0.7$, while their present-day values remain near the quintessence regime. Here, $W<-1/3$ shows the dark energy phase of the
universe. Finally, $W$ reaches $-1$ as $z\sim -1$, i.e., the universe goes to the $\Lambda$CDM stages due to the contributions
of VMCJG and VMCAG. However, their high-redshift behaviors differ markedly: the VMCJG model evolves toward a deep phantom regime, reaching $w_{DE}\sim-2$ to $-2.6$ at $z=2$, whereas the VMCAG model evolves toward $w_{DE}\approx0$, with slightly positive values for some datasets. Thus, the Abel parametrization exhibits a more physically conventional early-time behavior, while the strong high-redshift phantom evolution of the Jacobi model may be difficult to reconcile with the expected matter-dominated early universe. This divergence at $z\gtrsim1$ provides an important discriminating feature that can be tested with future high-redshift observations.

$\bullet{}$ $W$-$W'$ plane:\\

The $w$--$w'$ phase-plane trajectories in Fig.~\ref{omega} reveal distinctly different dynamical behaviors depending on the diagnostic quantity used.

In the first case, all four dataset combinations approach the $\Lambda$CDM fixed point $(-1,0)$ and then move into the phantom region ($w_{\rm DE} < -1$), following a pronounced freezing-type path ($w'_{\rm DE} < 0$) down to a minimum of about $-1.5$ to $-1.65$ near $w_{\rm DE} \sim -1.7$ to $-2.0$, before turning back toward less negative $w'$ as $w$ becomes strongly phantom ($\lesssim -2.5$) \cite{jawad2017cosmological, das2025f}. Among these, the DESI DR2 trajectory shows the most extreme excursion, while the CC+BAO curve stays closest to $\Lambda$CDM, and the DESI DR2 + DES-SN5YR and DESI DR2 + Union3 paths nearly overlap throughout.

In contrast, the second case shows all trajectories confined to $w' > 0$, indicating pure thawing behavior with no phantom crossing. Here $w_{\rm DE}$ ranges only from about $-0.9$ to $+0.1$, staying well away from the $\Lambda$CDM point. The curves converge to a common minimum of $w' \approx 0.14$ near $w \approx -0.6$, then diverge again, with the DESI DR2 + DES-SN5YR combination reaching the highest $w'$ values at both ends.

Overall, the two panels demonstrate that whether the reconstructed dark-energy dynamics allows phantom crossing and freezing evolution or remains confined to standard thawing quintessence depends critically on the reconstruction quantity used. The choice of dataset primarily influences the amplitude of the features, not their qualitative nature.

\begin{figure}[ht]
    \centering
    \begin{subfigure}[b]{0.48\textwidth}
\includegraphics[width=\textwidth]{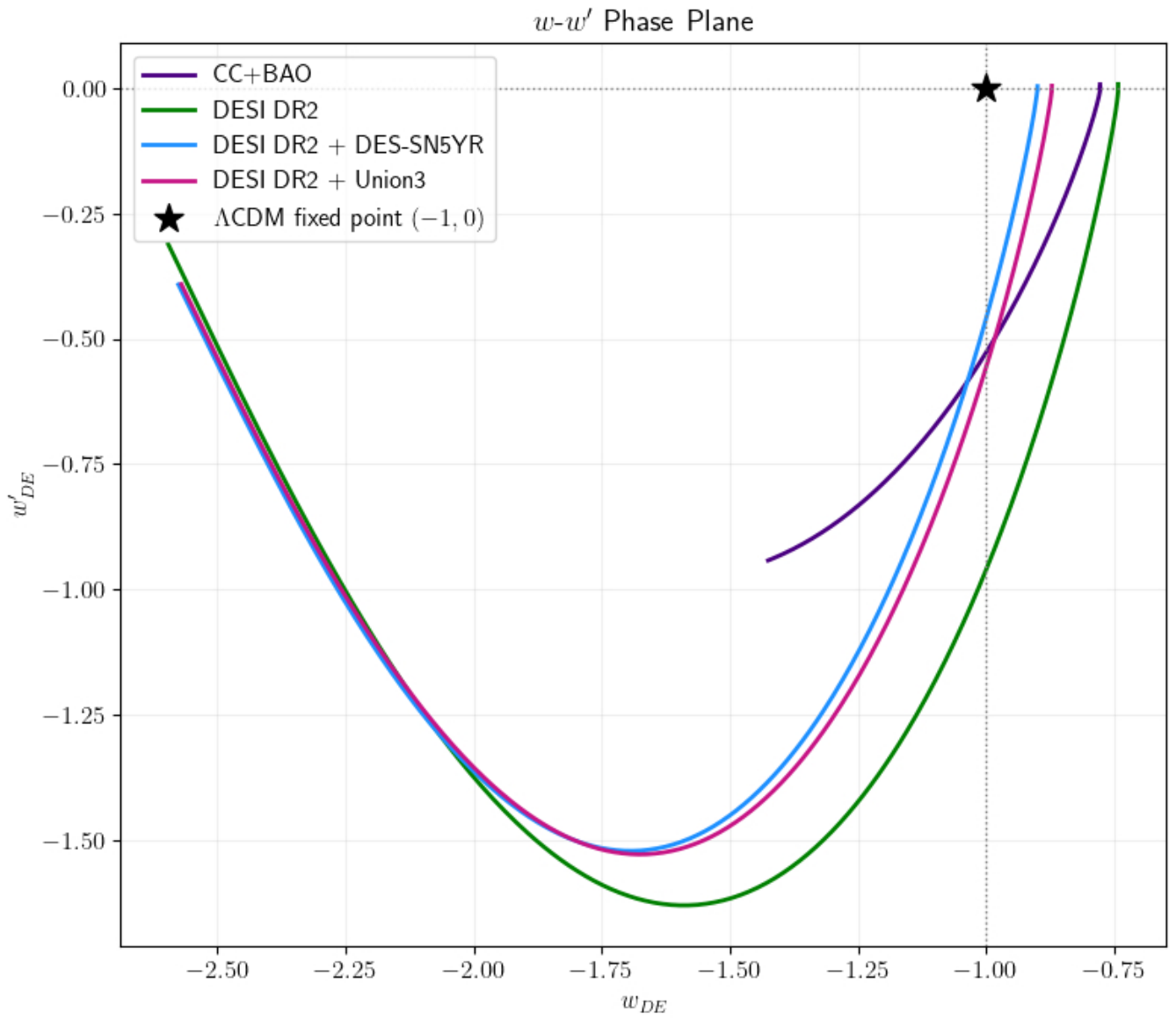}
        \caption{}
    \end{subfigure}
    \begin{subfigure}[b]{0.48\textwidth}
\includegraphics[width=\textwidth]{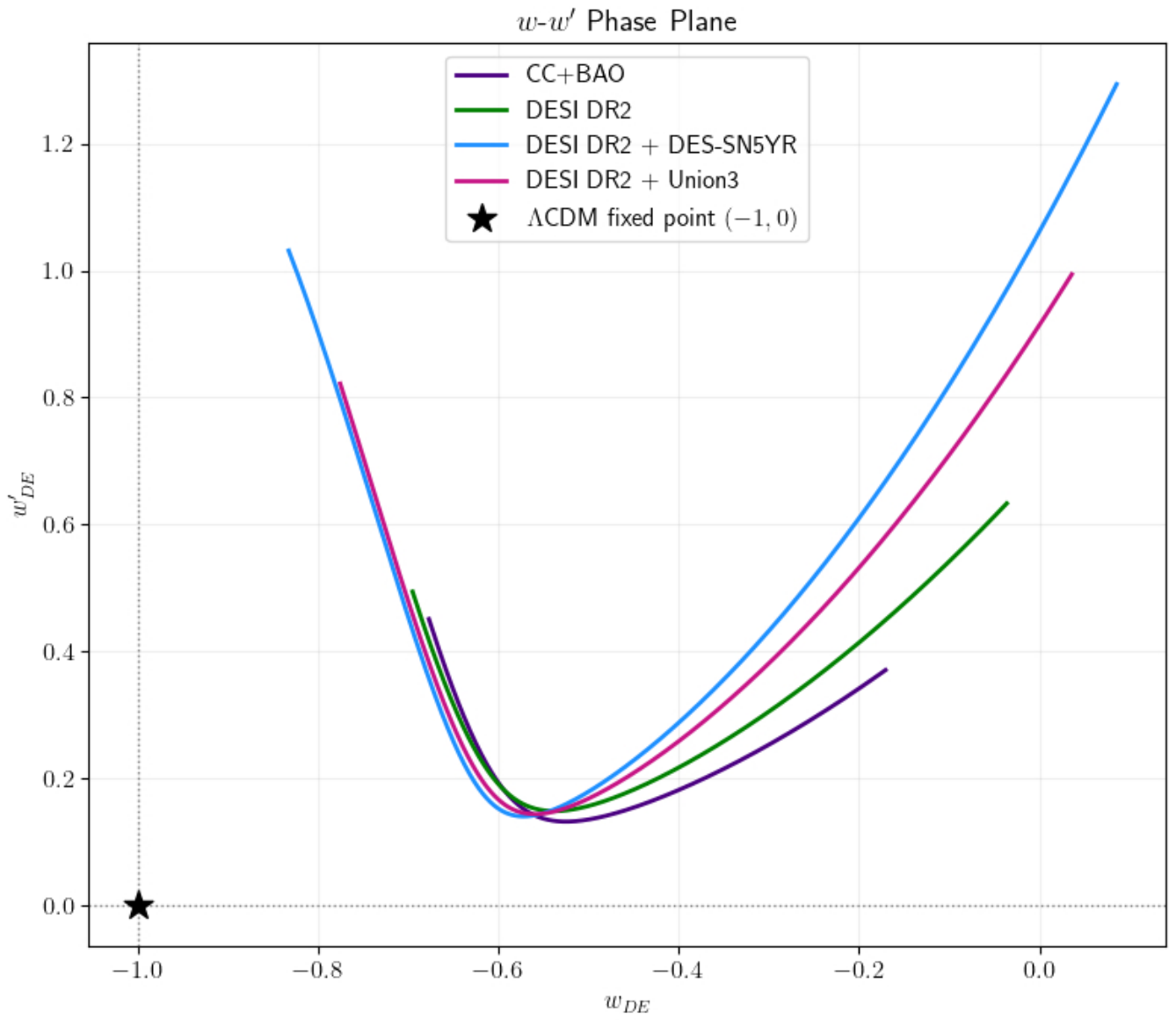}
        \caption{}
    \end{subfigure}
    \caption{Evolution of $\omega-\omega'$ analysis for the
(a) Variable Modified Chaplygin-Jacobi Gas (b) Variable Modified
Chaplygin-Abel Gas models for the considered observational datasets.}
    \label{omega}
\end{figure}

$\bullet{}$ Deceleration parameter:\\
\begin{figure}[ht]
    \centering
    \begin{subfigure}[b]{0.48\textwidth}
        \includegraphics[width=\textwidth]{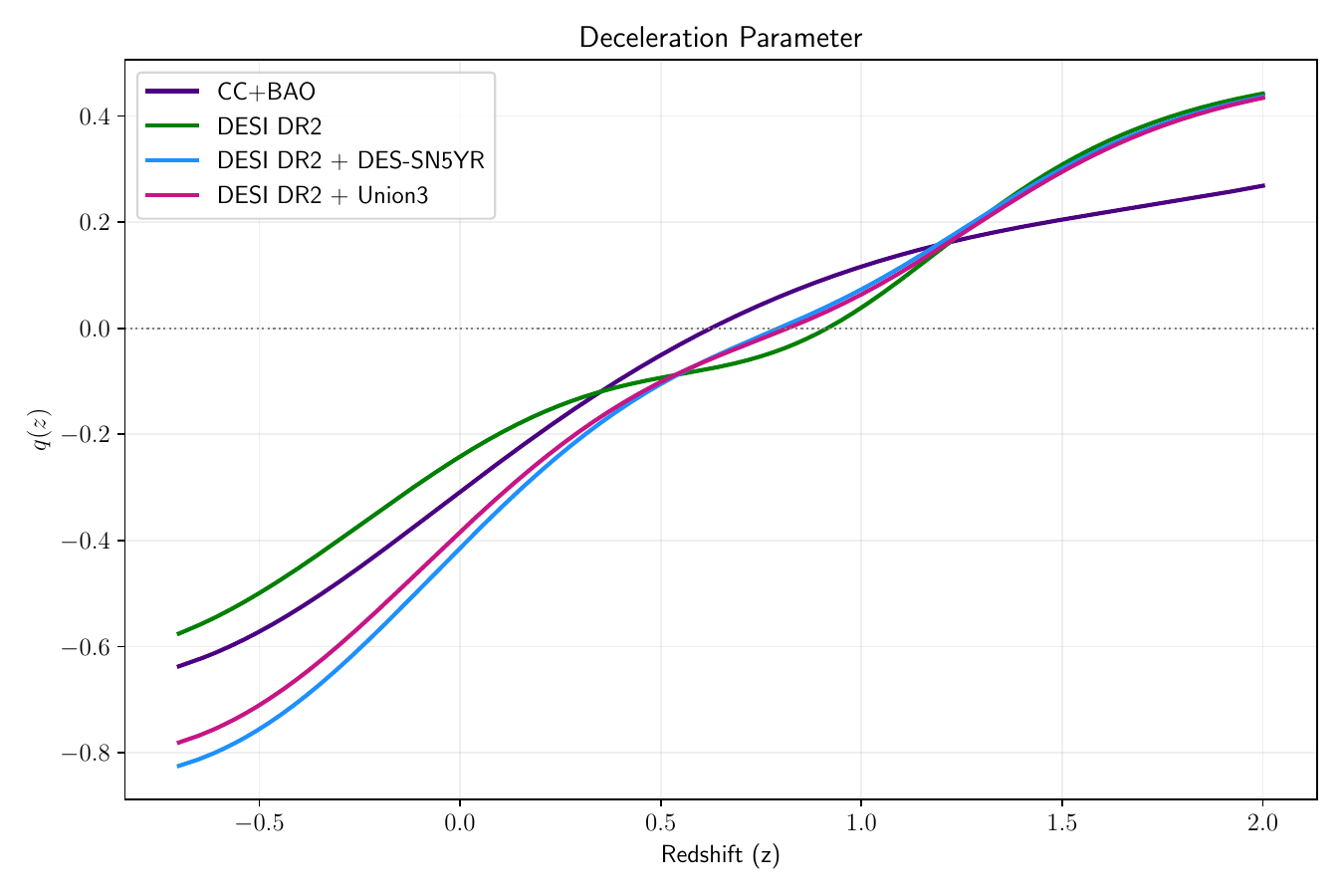}
        \caption{}
    \end{subfigure}
    \begin{subfigure}[b]{0.48\textwidth}
        \includegraphics[width=\textwidth]{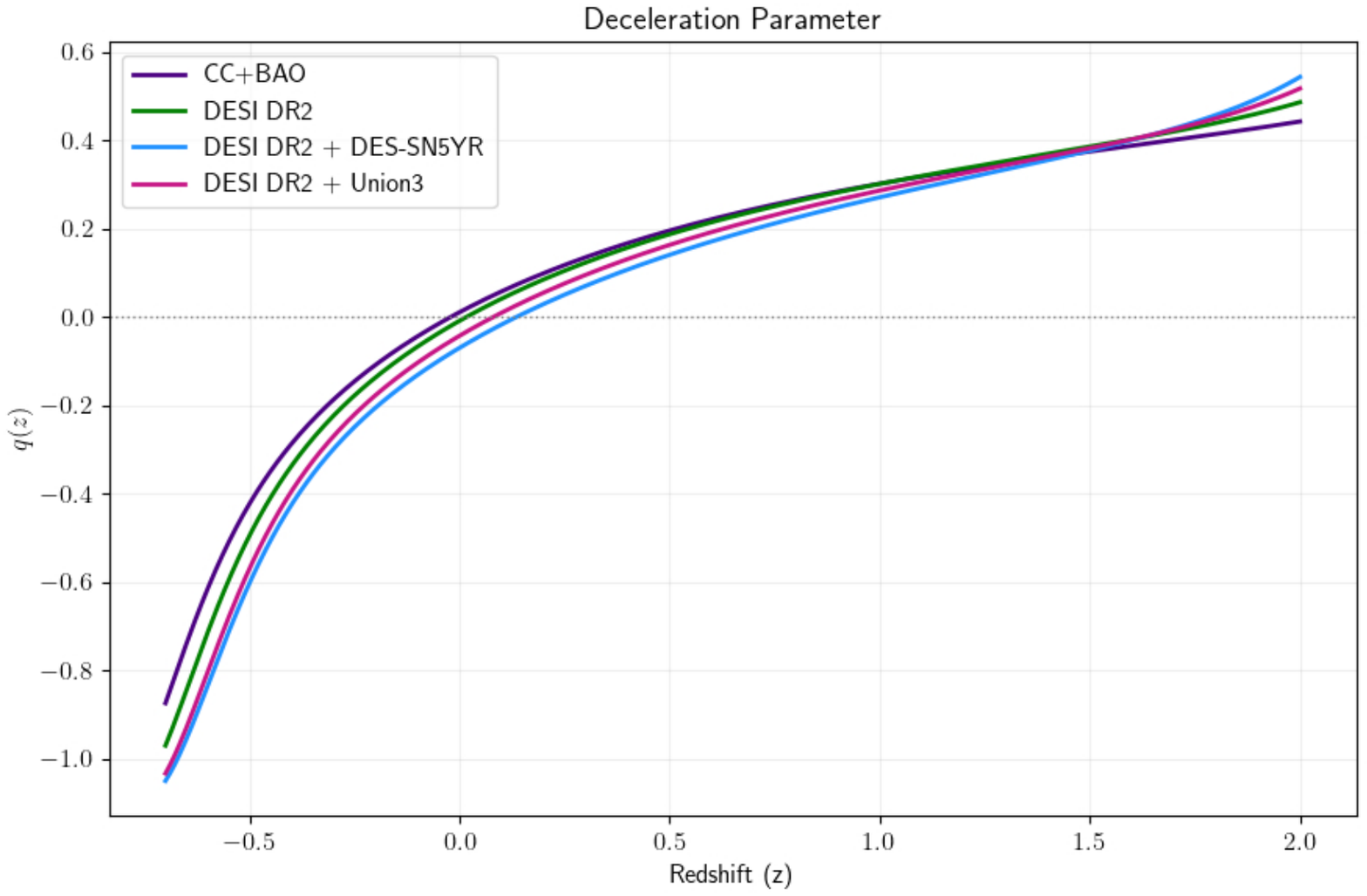}
        \caption{}
    \end{subfigure}
    \caption{Evolution of the deceleration parameter $q(z)$ for the
(a) Variable Modified Chaplygin-Jacobi Gas (b) Variable Modified
Chaplygin-Abel Gas models for the considered observational datasets.}
    \label{dece}
\end{figure}

In order to discuss the deceleration or acceleration phase of the universe, we
need to study the deceleration parameter, which is given as
$q=-\frac{\ddot{a}}{aH^{2}}$. For the deceleration phase of the
universe, $q>0$, and for the acceleration phase of the universe,
$q<0$ \cite{aviles2012cosmography,capozziello2011comprehensive, capozziello2013cosmographic}. \\
 From Fig. \ref{dece}, we can see that both models show the expected transition from a decelerating matter-dominated phase at high redshift ($q>0$) to late-time acceleration ($q<0$), with a broadly consistent transition redshift of $z_t\sim0.6$-$1.0$. The main difference lies in the strength of acceleration: the VMCAG model reaches more negative values, with $q\approx-1.1$ for the SN-inclusive datasets, briefly entering the phantom-like regime ($q<-1$), while the VMCJG model remains moderate, with $q\gtrsim-0.85$. At high redshift, both models approach $q\sim0.3$-$0.6$, close to the matter-dominated value $q=0.5$. The DESI DR2-only Jacobi fit shows a non-monotonic feature with a double zero-crossing around $z\sim0.3$-$0.9$, likely reflecting the weaker constraints on $(\alpha,\beta)$ from BAO data alone, consistent with the broader posterior contours discussed earlier.

$\bullet{}$ Cosmographic parameters:\\

There are another parameters named as cosmographical parameters
for understanding the past and future evolution of the universe \cite{AlMamon:2015nzh, Debnath:2019isy, Cai:2010uf}.
The cosmography is the study of the scale factor by expanding it by
Taylor series with respect to the cosmic time. The cosmographical
parameter like jerk ($J$), snap ($S$) parameters
are \cite{Mandal:2020buf}
\begin{eqnarray}
J=\frac{\dddot{a}}{aH^{3}}=(1+z)\frac{dq}{dz}+q(1+2q),\\
S=\frac{a^{(4)}}{aH^{4}}=-(1+z)\frac{dJ}{dz}-J(2+3q),\\
\end{eqnarray}
So the cosmographical parameters are higher-order derivatives of
the deceleration parameter $q$. \\

\begin{figure}[ht]
    \centering
    \begin{subfigure}[b]{0.48\textwidth}
\includegraphics[width=\textwidth]{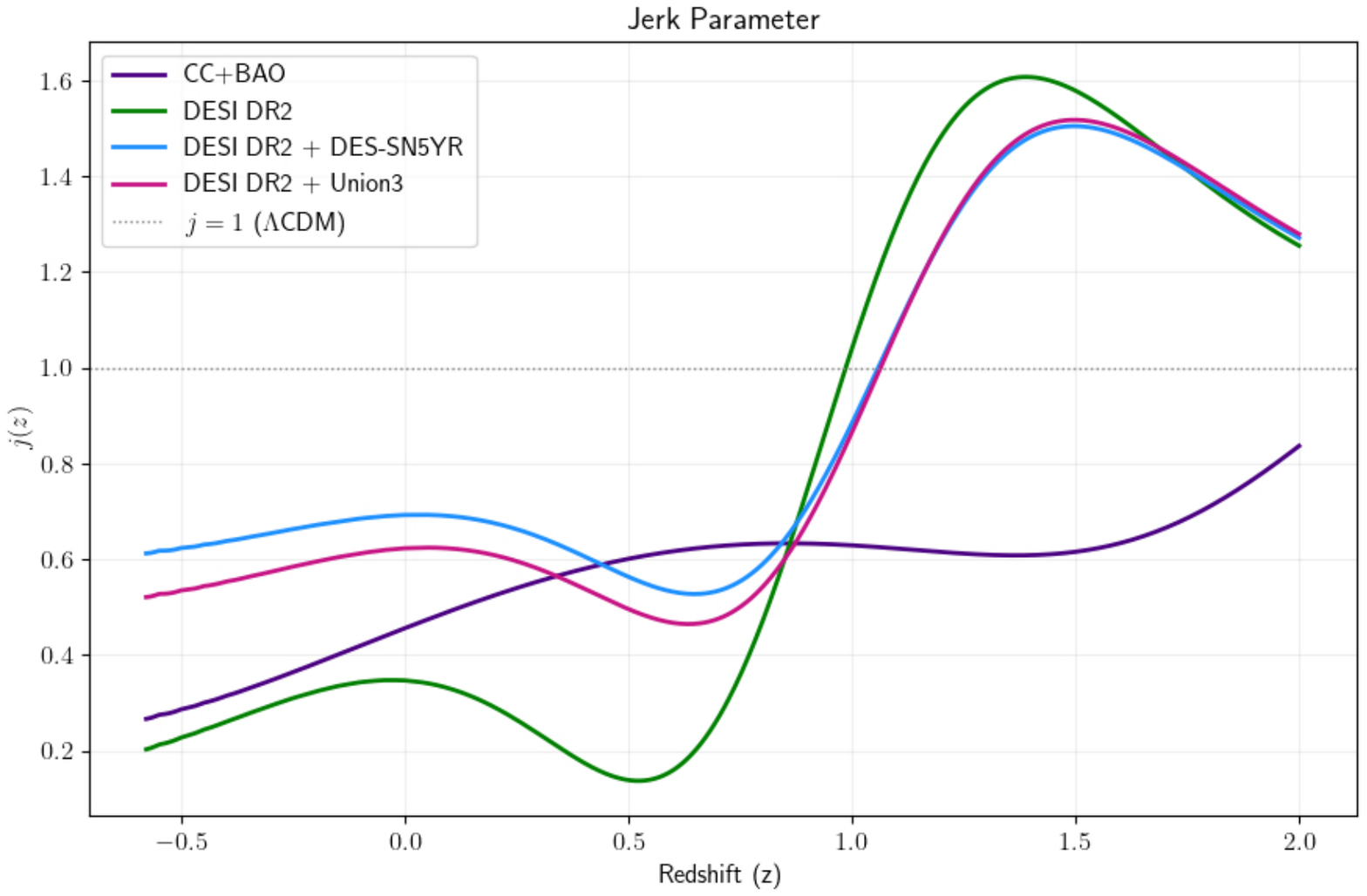}
        \caption{}
    \end{subfigure}
    \begin{subfigure}[b]{0.48\textwidth}
\includegraphics[width=\textwidth]{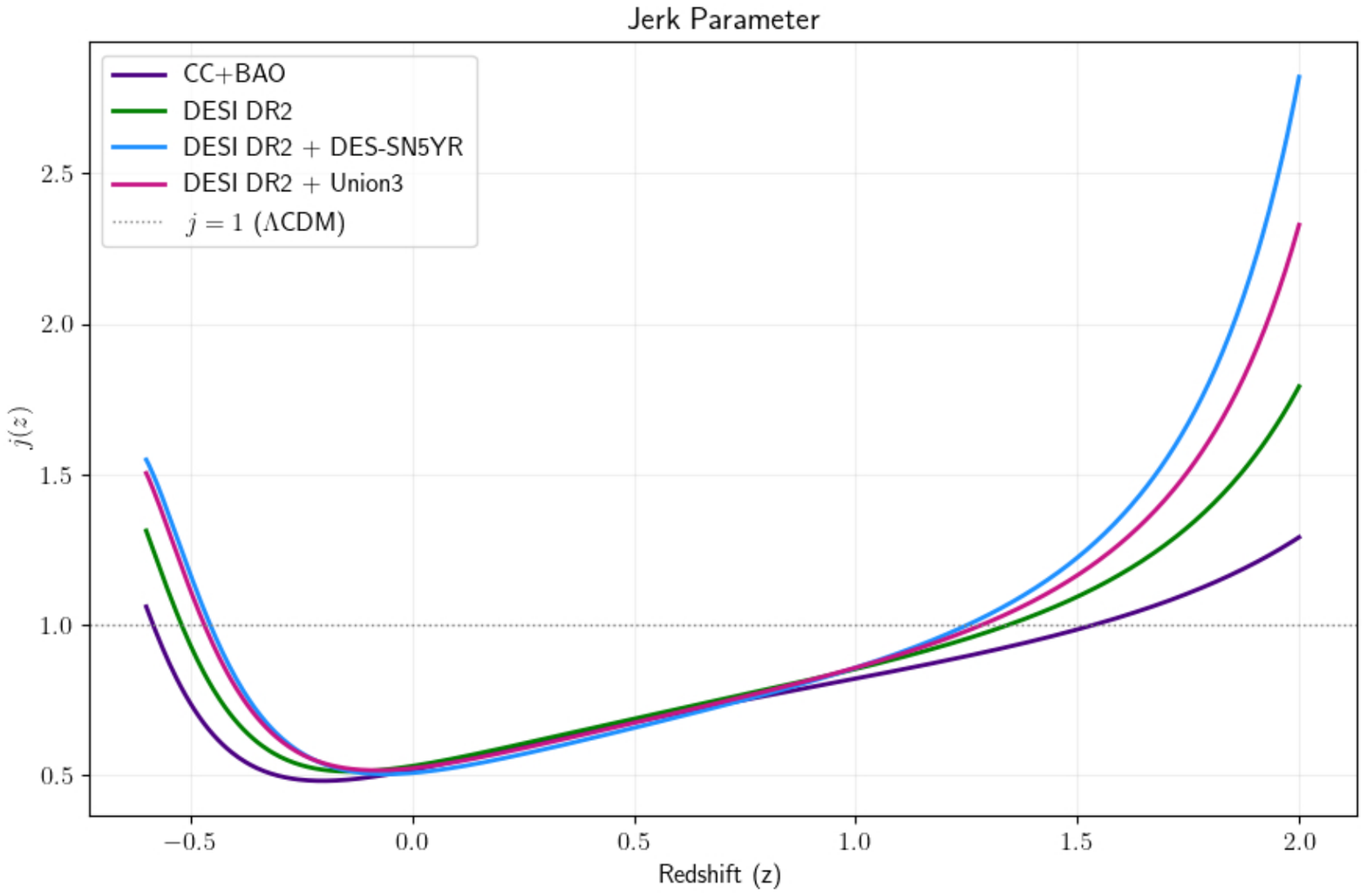}
        \caption{}
    \end{subfigure}
    \caption{Evolution of the jerk parameter $j(z)$ for the
(a) Variable Modified Chaplygin-Jacobi Gas (b) Variable Modified
Chaplygin-Abel Gas models for the considered observational datasets.}
    \label{jerk}
\end{figure}

Fig. \ref{jerk} demonstrates that the jerk parameter $j(z)$ provides an additional distinction between the two parametrizations. The VMCJG model exhibits an oscillatory evolution around the $\Lambda$CDM value $j=1$, remaining mostly below unity at low redshift, reaching $j\sim1.3$-$1.7$ near $z\sim1.2$-$1.5$, and approaching $j\sim1.25$-$1.3$ by $z=2$. The DESI DR2-only case shows a pronounced dip near $j\approx0.5$, reflecting the weaker constraints from BAO data alone. In contrast, the VMCJG model follows a smoother evolution: after a minimum of $j\sim0.2$-$0.3$ near $z\sim0$-$0.2$, it rises above $j=1$ around $z\sim1$-$1.3$ and reaches $j\sim3$-$5.4$ at $z=2$. Hence, although the VMCAG model exhibits a more conventional $j(z)$ evolution, it shows a substantially stronger high-redshift deviation from $\Lambda$CDM in higher-order kinematics, while the VMCJG model displays a more oscillatory and dataset-dependent behavior.\\
\begin{figure}[ht]
    \centering
    \begin{subfigure}[b]{0.48\textwidth}
\includegraphics[width=\textwidth]{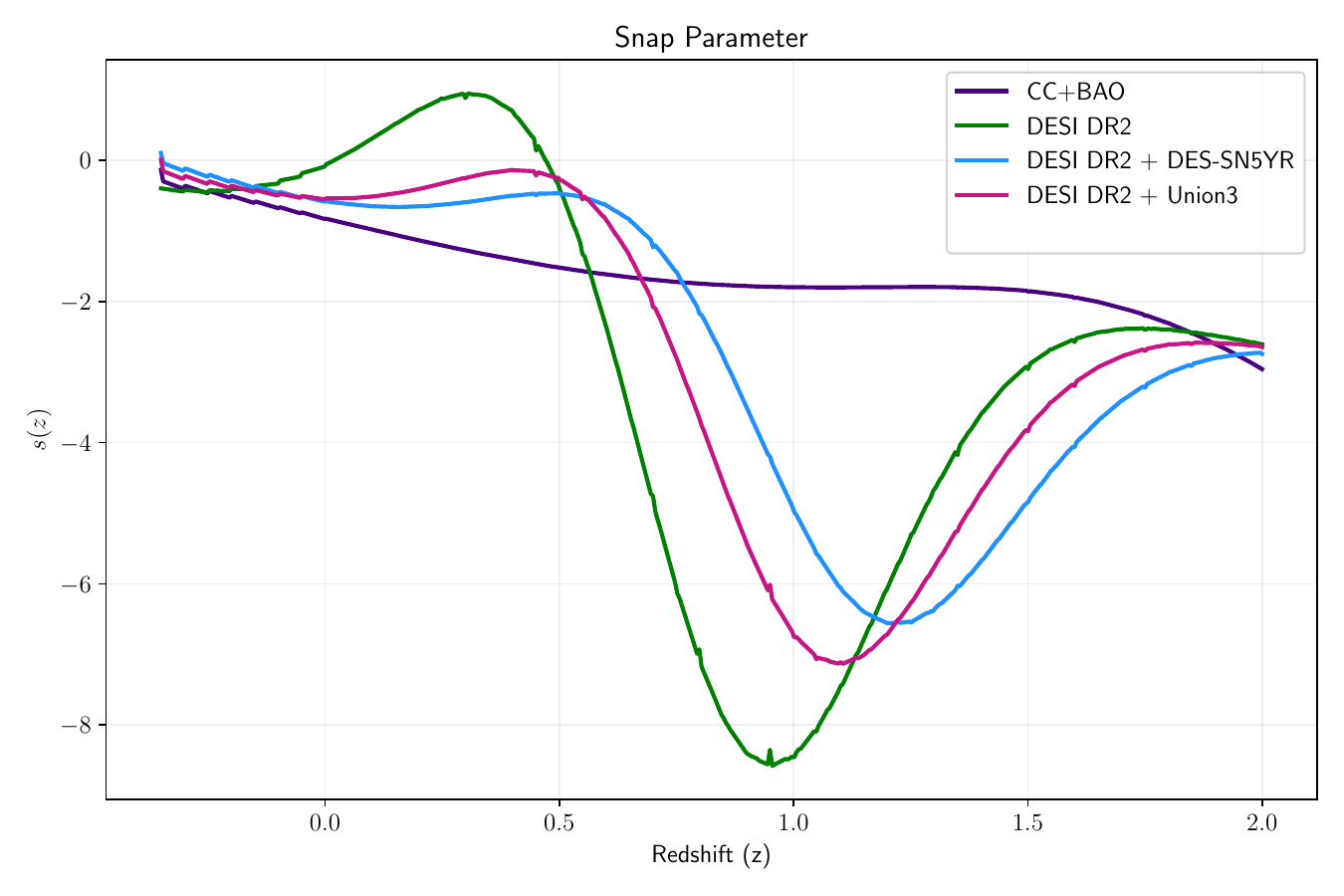}
        \caption{}
    \end{subfigure}
    \begin{subfigure}[b]{0.48\textwidth}
\includegraphics[width=\textwidth]{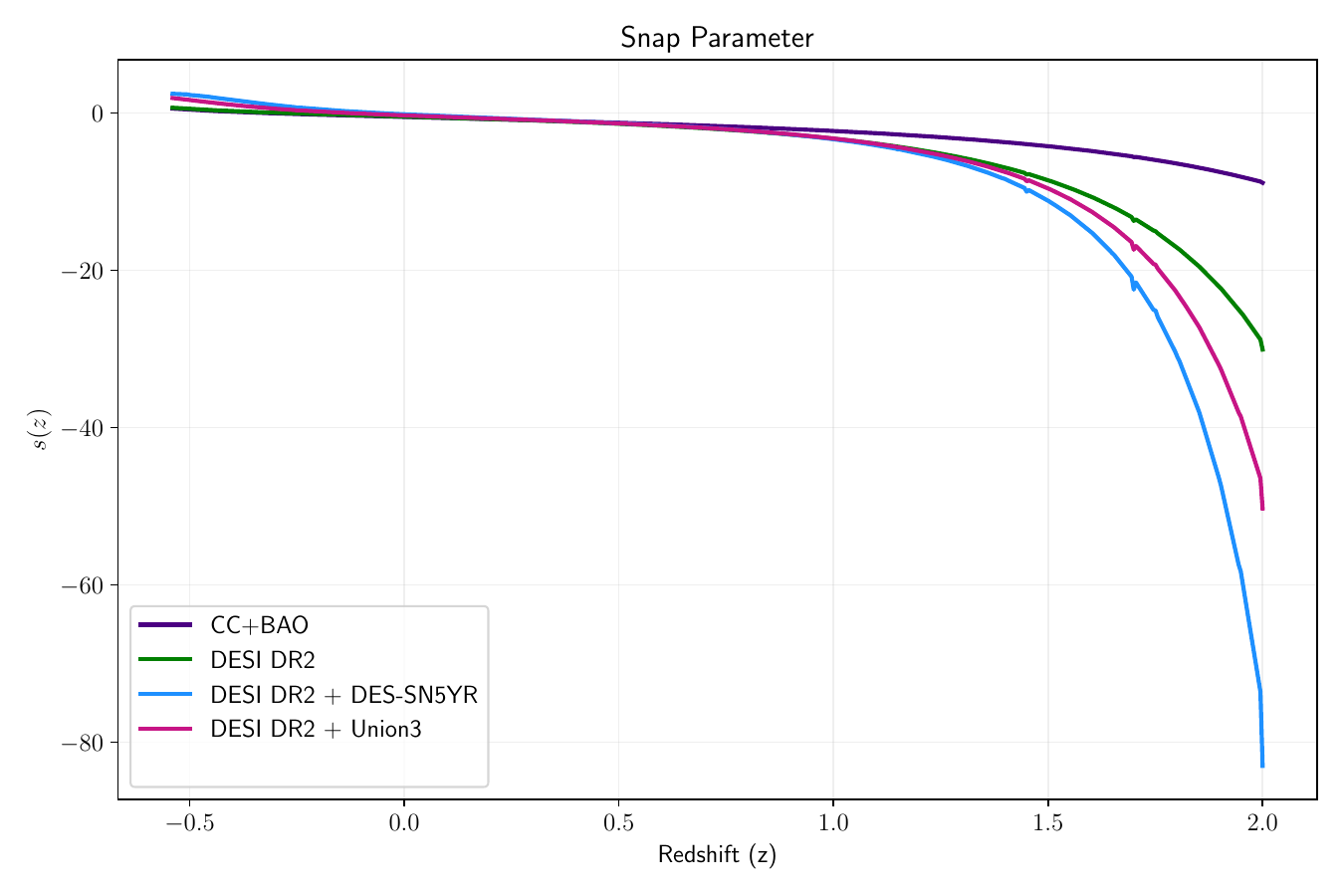}
        \caption{}
    \end{subfigure}
    \caption{Evolution of the snap parameter $s(z)$ for the
(a) Variable Modified Chaplygin-Jacobi Gas (b) Variable Modified
Chaplygin-Abel Gas models for the considered observational datasets.}
    \label{snap}
\end{figure}
The snap parameter $s(z)$ further differentiates the two parametrizations at fourth order, as evaluated from Fig.\ref{snap}. The VMCJG model shows a strongly non-monotonic evolution, with a pronounced minimum of $s\sim-1.8$ to $-8.5$ around $z\sim0.7$-$1.0$, followed by partial recovery toward $z=2$. Additional extrema in the DESI DR2-only reflect the dataset-dependent features seen in the lower-order diagnostics. In contrast, the VMCAG model remains relatively smooth, staying near zero at $z\lesssim1$, but then declines sharply to very large negative values, reaching $s\sim-9$ to $-83$ by $z=2$. Thus, the VMCJG model exhibits intermediate-redshift oscillatory behavior, whereas the Abel model develops a much stronger high-redshift divergence, indicating a substantially different departure from $\Lambda$CDM despite its more conventional $w_{DE}(z)$ evolution.\\
\begin{figure}[ht]
    \centering
    \begin{subfigure}[b]{0.48\textwidth}
\includegraphics[width=\textwidth]{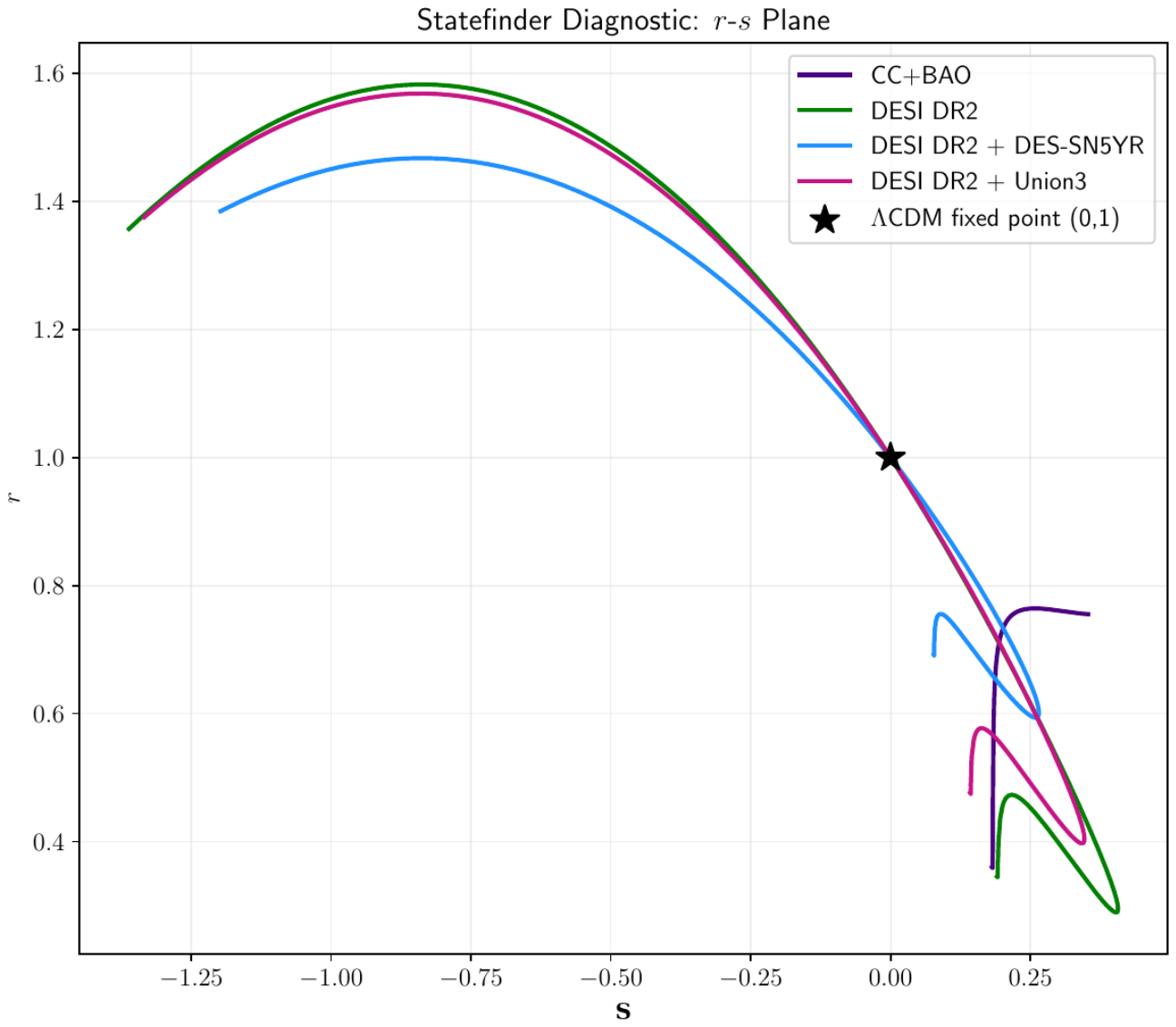}
        \caption{}
    \end{subfigure}
    \begin{subfigure}[b]{0.48\textwidth}
\includegraphics[width=\textwidth]{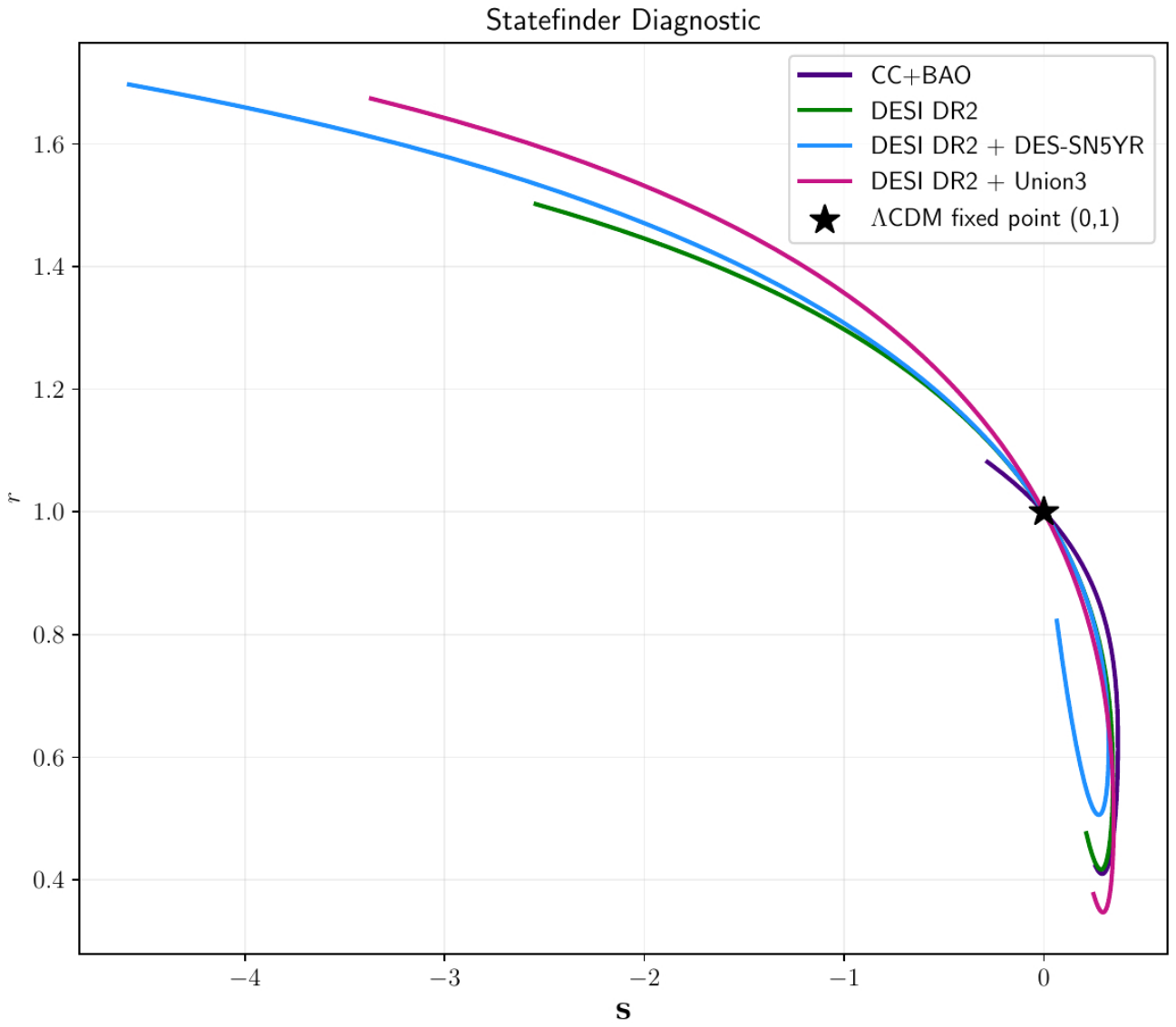}
    \caption{}
    \end{subfigure}
    \caption{Evolution of the statefinder parameter $r-s$ for the
(a) Variable Modified Chaplygin-Jacobi Gas (b) Variable Modified
Chaplygin-Abel Gas models for the considered observational datasets.}
    \label{state}
\end{figure}
$\bullet{\textbf{Statefinder Diagnostic}}$:
The statefinder diagnostic provides a useful geometrical tool for distinguishing
different dark-energy models through higher-order kinematic quantities \cite{sahni2003statefinder, zhang2005statefinder, ghaffari2015statefinder}. It is defined by the dimensionless parameters $(r,\textbf{S})$, where the statefinder pair is given by
\begin{equation}
    r \equiv \frac{\dddot{a}}{aH^3},
    \qquad
    s \equiv \frac{r-1}{3\left(q-\frac{1}{2}\right)},
    \label{eq:statefinder}
\end{equation}
where $q$ is the deceleration parameter and an overdot denotes differentiation w.r.t. cosmic time. The $\Lambda$CDM model corresponds to the fixed
point $(r,s)=(1,0)$, making the statefinder plane particularly useful for quantifying deviations from the concordance model.

From Fig. \ref{state}, in both \(r\)-\(\textbf{S}\) plane diagrams, all four dataset combinations pass through or near the \(\Lambda\)CDM fixed point \((\textbf{S},r)=(0,1)\), confirming consistency with the standard model at some epoch. From there, the curves bend into the upper-left region (\(\textbf{S}<0,\ r>1\)), associated with Chaplygin gas-like behavior, then loop back through the fixed point and continue into the lower-right quadrant (\(\textbf{S}>0,\ r<1\)), characteristic of quintessence. This indicates that each trajectory samples both regimes over its evolution—early phases resemble Chaplygin gas, while later stages shift toward quintessence.

The DESI DR2 and DESI DR2+Union3 curves show the most pronounced curvature, peaking at \(r \approx 1.55\)-\(1.6\), whereas the DESI DR2+DES-SN5YR combination remains more damped, with a peak around \(r \approx 1.35\)-\(1.47\). All trajectories also develop a small loop or kink well into the quintessence region (\(\textbf{S} \approx 0.1\)-\(0.4,\ r \approx 0.3\)-\(0.8\)), indicating rapid, non-monotonic changes in the deceleration parameter's derivatives at late times; the CC+BAO curve extends furthest in \(\textbf{S}\) (\(\approx 0.4\)) with the least pronounced loop.

Despite these differences in amplitude and loop extent, the qualitative agreement across all datasets supports a dynamical dark energy history that transitions between Chaplygin-gas like and quintessence-like phases near the present epoch, rather than a static \(\Lambda\)CDM cosmology.\\

$\bullet{}$ \textbf{Stability}:\\

\begin{figure}[ht]
    \centering
    \begin{subfigure}[b]{0.48\textwidth}
\includegraphics[width=\textwidth]{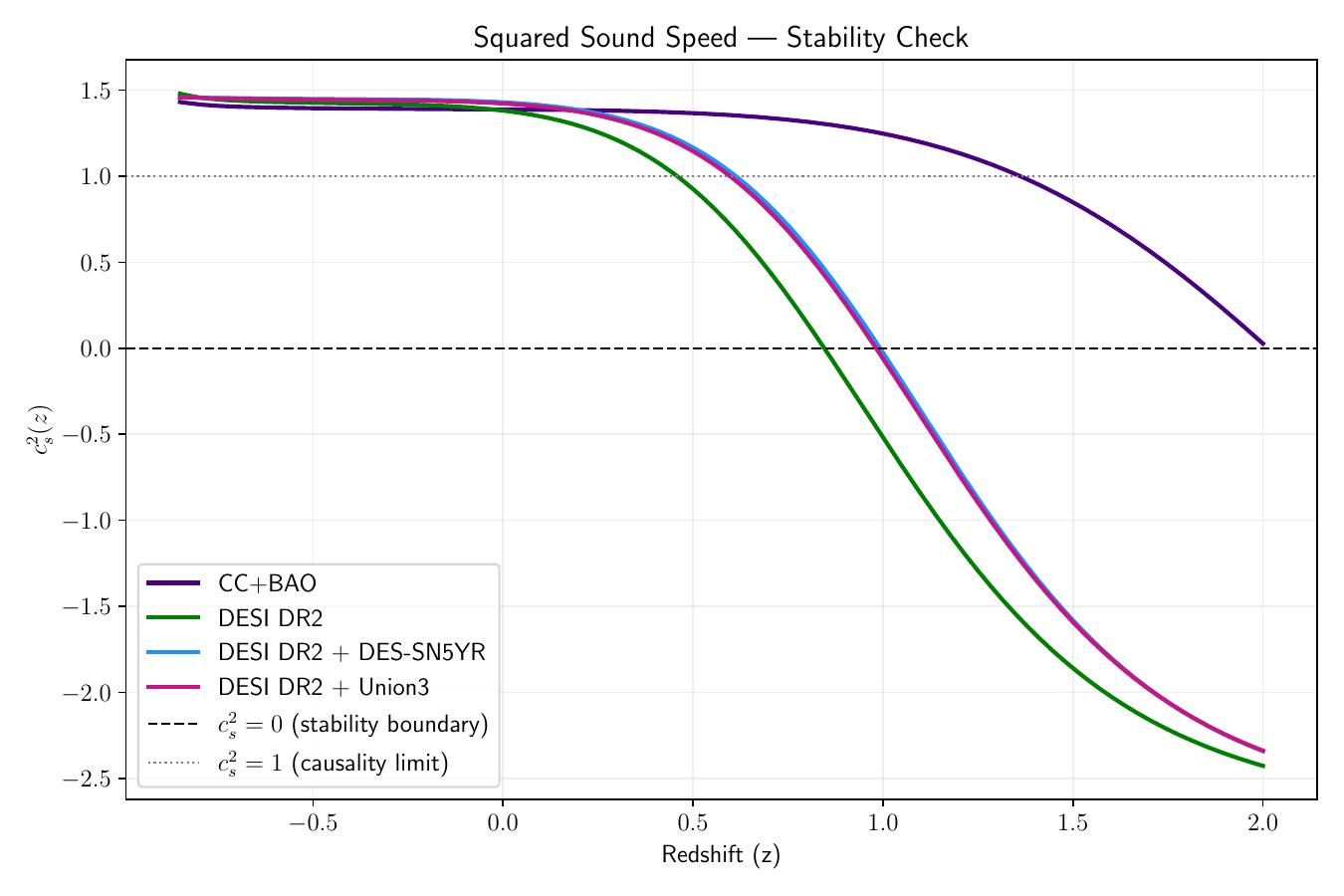}
        \caption{}
    \end{subfigure}
    \begin{subfigure}[b]{0.48\textwidth}
\includegraphics[width=\textwidth]{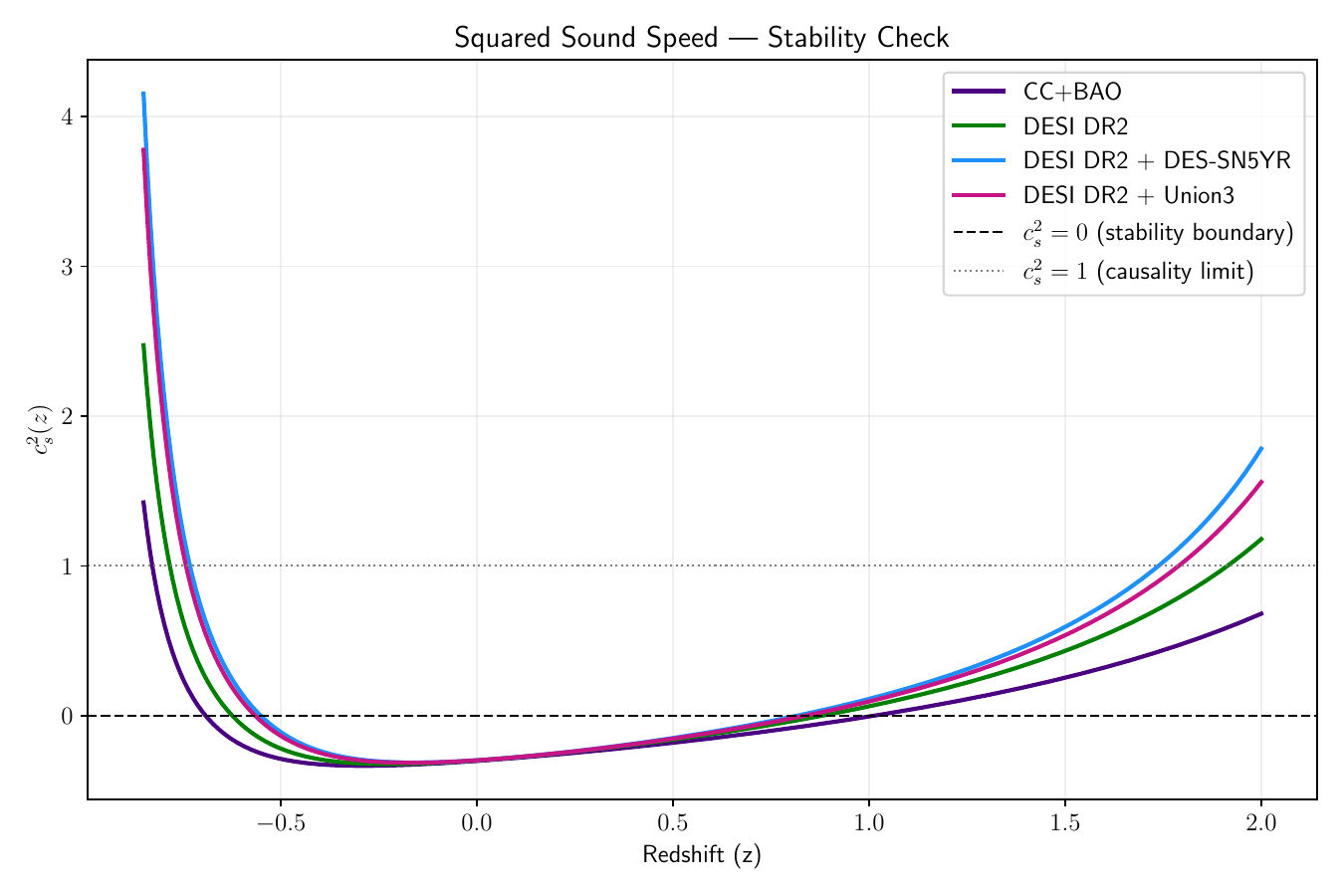}
        \caption{}
    \end{subfigure}
    \caption{Evolution of the squared sound speed $c_s^2(z)$ for the
(a)Variable Modified Chaplygin-Jacobi Gas (b) Variable Modified
Chaplygin-Abel Gas models for the considered observational datasets,
illustrating their perturbative stability and causal behavior.}
    \label{svs}
\end{figure}
To study the stability against small perturbations, we can write
the squared speed of sound in the form
$c_{s}^{2}=\frac{dp}{d\rho}$. If $c_{s}^{2}>0$, we can say that
the model is classically stable while $c_{s}^{2}<0$ represents a classically unstable.\\

The squared sound speed $c_s^2(z)$ provides a crucial test of the perturbative stability of the two models, as seen from Fig.\ref{svs}. The VMCJG model is generally stable at low redshift, with $c_s^2>0$, but develops a negative sound speed at higher redshift, reaching $c_s^2\sim-2$ to $-2.5$ by $z=2$ for most datasets. In contrast, the VMCAG model becomes unstable in the low-redshift regime, with $c_s^2\sim-0.3$ to $-0.4$ over approximately $z\sim-0.3$ to $0.6$--$1.0$, before recovering at higher redshift. Both models also exhibit superluminal values ($c_s^2>1$) at the extremes of the redshift range. Thus, the Jacobi parametrization appears more viable over the observationally constrained low-redshift regime, whereas the Abel model faces a stability issue at late times. The superluminal behavior in both cases remains an important theoretical caveat.

\section{\bf{Discussions and Conclusions}}\label{con}

In this work, we explore the potential applications of thermodynamic geometry to obtain new physically relevant insights into cosmology. In particular, the central idea is that fluids possessing distinctive geometric properties may also exhibit unique cosmological behavior. \\

We constrained the free parameters of both models using four complementary observational data sets: CC+BAO, DESI DR2, DESI DR2+DES5Y, and DESI DR2+Union3. Through a Markov Chain Monte Carlo likelihood analysis, we obtained posterior distributions for the model parameters. The reconstructed parameter values reveal a clear physical distinction between the two Chaplygin-gas realizations and their relation to the standard $\Lambda$CDM baseline. Across all four dataset combinations, both the VMCAG and VMCJG models recover a higher Hubble constant $H_0$ than $\Lambda$CDM, with the discrepancy being most pronounced for the DESI DR2 + DES-SN5YR combination. In this case, $\Lambda$CDM attains its lowest inferred value, $H_0 = 66.943^{+0.382}_{-0.491}$, whereas both Chaplygin models remain at $H_0 \gtrsim 71$. This behavior indicates that the additional degrees of freedom in the Chaplygin-gas equation of state absorb part of the low-$H_0$/high-$\Omega_{m0}$ pull that the DES-SN5YR compilation exerts on $\Lambda$CDM, thereby partially decoupling the inferred Hubble constant from the well-known $\Omega_{m0}$-$H_0$ anti-correlation driven by supernova calibration systematics. Consistent with this interpretation, the matter density parameter $\Omega_{m0}$ in the VMCJG scenario is systematically lower than the corresponding $\Lambda$CDM value for every dataset, with differences of approximately $0.03$-$0.08$. This suggests that a fraction of the total energy budget attributed to pressureless matter in $\Lambda$CDM is instead being reinterpreted as an emergent unified dark-fluid contribution within this model.\\
The information criteria analysis reveals a data-dependent preference for the Chaplygin models. For DESI DR2 alone, all three models are nearly degenerate. However, for DESI DR2 + DES-SN5YR, the VMCJG model is strongly favoured over $\Lambda$CDM ($\Delta$AIC = $-18.50$, $\Delta$DIC = $-17.95$), while the VMCAG model is disfavoured ($\Delta$AIC = $+3.34$). This indicates that the Jacobi parametrization effectively absorbs the low-$H_0$/high-$\Omega_{m0}$ tension introduced by the supernova data. For DESI DR2 + Union3, both Chaplygin models are favored, with the Jacobi model again performing best ($\Delta$AIC = $-6.24$). Overall, the VMCJG model emerges as the most competitive alternative to $\Lambda$CDM, though the preference is dataset-dependent and warrants confirmation with future surveys.\\

The cosmological analysis further distinguishes the two parametrizations. The equation-of-state parameter $w_{\rm DE}(z)$ shows that both models cross the phantom divide $w=-1$ around $z \sim 0.3$--$0.7$, with present-day values in the quintessence regime. However, their high-redshift behaviors differ markedly: the VMCJG model evolves towards a deep phantom regime reaching $w_{\rm DE} \sim -2$ to $-2.6$ at $z=2$, while the VMCJG behavior model evolves towards $w_{\rm DE} \approx 0$. The $w$-$w'$ phase-plane analysis confirms this distinction, with the VMCJG model exhibiting freezing-type behaviour ($w'<0$) with phantom crossing, whereas the VMCAG model remains confined to pure thawing quintessence ($w'>0$) with no phantom crossing. The deceleration parameter $q(z)$ shows that both models exhibit the expected transition from deceleration to acceleration at $z_t \sim 0.6$-$1.0$. The VMCAG model reaches more negative values ($q \approx -1.1$), briefly entering the phantom-like regime, while the Jacobi model remains moderate ($q \gtrsim -0.85$). The jerk $j(z)$ and snap $s(z)$ parameters further differentiate the two models: the VMCJG model exhibits oscillatory behaviour around $\Lambda$CDM with moderate high-redshift deviations, while the VMCAG model shows smoother evolution at low redshift but develops substantially stronger high-redshift divergences, reaching $j \sim 3$--$5.4$ and $s \sim -9$ to $-83$ at $z=2$. The statefinder $(r,s)$ diagnostic confirms these trends, with the VMCJG model remaining closer to the $\Lambda$CDM fixed point while the VMCAG model exhibits larger departures. The statefinder diagnostic reveals that both models trace out trajectories that pass through the $\Lambda$CDM fixed point $(s,r)=(0,1)$ and sample both Chaplygin gas-like ($s<0$, $r>1$) and quintessence-like ($s>0$, $r<1$) regimes over their evolution, indicating a dynamical transition rather than a static cosmological constant. The DESI DR2 and DESI DR2+Union3 combinations exhibit the most pronounced curvature, peaking at $r \sim 1.55$--$1.6$, while the DESI DR2+DES-SN5YR combination remains more damped. All trajectories develop small loops or kinks in the quintessence region, signaling rapid non-monotonic changes in the deceleration parameter's derivatives at late times. \\

The perturbative stability analysis, however, reveals important caveats: the Jacobi model is stable at low redshift but becomes unstable at high redshift ($c_s^2 < 0$ for $z \gtrsim 2$), while the Abel model exhibits low-redshift instabilities ($c_s^2 \sim -0.3$ to $-0.4$ for $z \sim -0.3$ to $0.6$-$1.0$) before recovering at higher redshift. Both models also show superluminal behavior ($c_s^2 > 1$) at extreme redshifts. These findings suggest that while the background dynamics of both Chaplygin parametrizations remain competitive with $\Lambda$CDM, their perturbative stability requires further investigation before they can be considered fully viable alternatives.\\

In conclusion, we have demonstrated that both Chaplygin-gas parametrizations derived from geometrothermodynamics offer viable dynamical alternatives to $\Lambda$CDM, with the VMCJG model emerging as the most competitive candidate, particularly for the DESI DR2 + DES-SN5YR dataset, and that the extra equation-of-state freedom in both unified dark-fluid descriptions absorbs much of the compilation dependent shift in $H_{0}$ and $\Omega_{m0}$ seen in the standard model — pointing to unified dark-fluid geometrothermodynamics as a physically motivated and observationally robust alternative to $\Lambda$CDM. However, the perturbative stability analysis reveals important theoretical considerations that point to future work. While the background dynamics are viable, the perturbative sector requires further investigation. Future work will focus on extending this analysis to include CMB data, constructing a full effective field theory description to ensure perturbative stability, and applying the GTD framework to other generalized dark-energy scenarios.

\section*{Acknowledgment}

AK is thankful to IIEST, Shibpur, India, for providing Institute Fellowship (SRF).

\section{Data Availability Statement}
Data sharing not applicable to this article as no datasets were generated or analysed during the current study.

%%%%%%%%%%%%%%%%%%%%%%%%%%%%%%%%%%%%%%%%%%%%%%%%%%%%%%%%%%%%%%%%%%%%%%%%%%%%%%%%%%%%%%%%%%%%%%%%%%%%%%%%%%%%%%%%%%%%%%%%%%%%%%%%%%%%%5

\bibliographystyle{apsrev4-1}
\bibliography{ref}
%%%%%%%%%%%%%%%%%%%%%%%%%%%%%%%%%%%%%%%%%%%%%%%%%%%%%%%%%%%%%%%%%%%%%%%%%%%%%%%%%%%%%%%%%%%%%%%%%%%%%%%%%%%%%%%%%%
\end{document}